\documentclass[prl,twocolumn,showpacs,superscriptaddress,floatfix]{revtex4}

\usepackage{amsmath}			
\usepackage{amssymb}			
\usepackage{graphicx}			
\usepackage{epic}
  \graphicspath{{./Eps/}}

\begin{document}

\title{Fermion nodes and nodal cells of noninteracting and interacting fermions
}

\author{Lubos \surname{Mitas}}
\affiliation{Department of Physics, North Carolina State University, Raleigh,
NC 27695}

\date{\today}

%
\begin{abstract}
A fermion node is subset of fermionic configurations for which a real wave 
function vanishes due to the antisymmetry and the node divides the configurations
space into compact nodal cells (domains). We analyze the properties of fermion 
nodes of fermionic ground state wave functions for a number of systems. For several
models we demonstrate that noninteracting spin-polarized fermions in dimension two 
and higher have closed-shell ground state wave functions with the minimal two nodal
cells for any system size and we formulate a theorem which sumarizes this result. 
The models include periodic fermion gas, fermions on the surface of a sphere, 
fermions in a box. We prove the same property for atomic states with up to $3d$ 
half-filled shells. Under rather general assumptions we then derive that the same 
is true for unpolarized systems with arbitrarily weak interactions using 
Bardeen-Cooper-Schrieffer (BCS) variational wave function. We further show that pair 
correlations included in the BCS wave function enable singlet pairs of particles 
to wind around the periodic box without crossing the node pointing towards the 
relationship of nodes to transport and many-body phases such as superconductivity.
Finally, we point out that the arguments extend also to fermionic temperature
dependent/imaginary-time density matrices. The results reveal fundamental properties 
of fermion nodal structures and provide new insights for accurate constructions of 
wave functions and density matrices in quantum and path integral Monte Carlo methods.
\end{abstract}

\pacs{02.70.Ss, 03.65.Ge}

\maketitle

%
\section{\label{sec:level1} Introduction}
Let us consider a system of fermions described 
by a real wave function $\Psi(R)$ where $R$ denotes 
fermions coordinates 
 $R=({\bf r}_1, ...,{\bf r}_N)$. Due to the antisymmetry, there
exists a subset of fermion configurations for which the 
wave function vanishes and
this subset is called the fermion node.
The fermion node can be  
implicitly defined by 
$\Psi(R)=0$, assuming that the nodal set does not include
configurations for which the wave function vanishes because
of other reasons, eg,
boundary conditions.
In general, for $N$  spin-polarized fermions in  
  a $d$-dimensional space,  the fermion node is
 a $(dN-1)-$dimensional manifold
(hypersurface). 
It is a well-known fact that for $d>1$ the
antisymmetry alone does not specify 
fermion nodes completely. This is not difficult to understand
since antisymmetry
fixes only lower-dimensional 
coincidence hyperplanes with dimensionalities $(dN-d)$.
Therefore the fermion nodes and their properties are determined   
by interactions and many-body effects.

The fermion nodes play an important role in 
 quantum Monte Carlo (QMC) methods which belong to the
 most promising and productive
approaches for studying quantum many-body
systems.
Let us briefly describe the basic idea of QMC and its relationship 
to fermion nodes. 
Consider a  Hamiltonian $H$ and  a
trial function $\Psi_T(R)$ which approximates the ground state
$\Psi_0(R)$ of $H$ within a given
symmetry sector. It is straightforward to show that 
$\lim_{\tau\to\infty}\exp(-\tau H)\Psi_T\propto \Psi_0$ 
where $\tau$ is a real parameter 
(imaginary time). This imaginary time projection can be conveniently
carried out by simulating a stochastic process 
which maps onto 
the imaginary time  Schr\"odinger equation.   
The wave function is represented by an
ensemble  of $R$-space sampling points which are propagated
 according to  $G(R,R', \tau)=
\langle R |\exp(-\tau H)|R'\rangle$ and for large $\tau$
the ensemble becomes distributed according to $\Psi_0$.
Unfortunately, the straightforward application of this idea
to fermionic systems encounters a fundamental complication
in the fermion sign problem \cite{qmchistory, hammond, jbanderson}.
The fermion sign problem makes QMC studies of large fermionic systems
difficult since the statistical errors 
of fermionic expectation values grow exponentially in the projection
time $\tau$ and also in the number of particles $N$. 

 One possibility for avoiding the fermion sign problem
is to employ the fixed-node approximation 
which restricts the fermion node
 of the solution $\Psi_0(R)$
to be identical to the fermion
node of an appropriate trial function $\Psi_T(R)$.
 The fixed-node approximation introduces an energy bias which
scales as the square of the difference between the exact and approximate
nodes and therefore the accuracy of 
fermion nodes becomes very important. In particular,
for the exact node one can obtain the exact energy in computational
time proportional to a low-order polynomial in $N$. The fixed-node
approximation has been very successful even for rather approximate
nodes of 
commonly used trial wave functions which are based on
Hartree-Fock (HF) determinants or a few-determinant post-HF expansions.
The fixed-node QMC electronic structure calculations
using HF or post-HF nodes usually recover
between 90 and 95 \%
of the correlation energy 
and energy differences 
agree with experiments typically within a few percent.  
Such encouraging results have been observed across many 
systems such as atoms, molecules, clusters and solids \cite{qmcrev}.
It is remarkable that
QMC methods have enabled us to "corner" the correlation energy problem
into the last few percent of the correlation energy using algorithms which 
polynomial scaling.




Unfortunately,
for many problems even a few percent of the correlation energy can be 
significant. Typical
examples are  transition metal systems where the fixed-node error
can be of the order of several eVs. 
It is therefore clear that better understanding of fermion nodes
could be an important step forward for the QMC methodology and beyond.
 In addition, the fermion nodes are
related to other physical quantities and could shed light on other many-body phenomena 
which currently are not completely understood.

In order to introduce the basic properties of the fermion nodes 
and to illustrate the problems 
involved we will first mention a few unique cases of interacting systems
for which the nodes are known exactly.
These examples include a few two- and three-electron atomic states, namely,
triplets of He atom  $^3S(1s2s)$, $^3P(2p^2)$
\cite{dariohe}
and the exact node of a three-electron atomic state
   $^4S(p^3)$ \cite{bajdichnode}. 
 The wave functions of these
 high symmetry states can be parametrized by appropriate 
coordinate  maps in which the node is described
by a single variable. For example, in the case of He triplet $^3S(1s2s)$ state the 
corresponding "nodal coordinate" is 
$\cos\beta={\bf r}^+_{12}\cdot {\bf r}_{12}/( r^+_{12} r_{12})$ where
${\bf r}^+_{ij}={\bf r}_{i}+{\bf r}_{j} $, ${\bf r}_{ij}={\bf r}_{i}-{\bf r}_{j} $.
For the triplet $^3P(2p^2)$ the relevant variable is
$\cos\omega'={\bf z}_0\cdot ({\bf r}_{1}\times {\bf r}_{2})$ assuming 
that the
$P$ state is oriented along the $z$-axis which is specified by
the unit vector ${\bf z}_0$ \cite{dariohe,bajdichnode}.
For the quartet state 
 $^4S(p^3)$
the node is captured by the variable 
$\cos\omega={\bf r}_{1}\cdot ({\bf r}_{2}\times {\bf r}_{3})$.
The node in these 
systems is encountered whenever the corresponding variable,
 $\cos\beta,\cos\omega'$
or $\cos\omega$,
vanishes.

The nodal surface divides the space of fermion configurations into
nodal cells, sometimes also called domains.
(The name nodal cell was introduced in the previous paper \cite{davidnode}.
In algebraic geometry and topology
 nodal cells are usually called nodal domains, see, for example,
Ref. \cite{berger}. We will use
both of these two expressions interchangeably.)
 The few known examples of exact nodes mentioned in the preceding paragraph help to illustrate an 
important point, which has been conjectured for the fermionic ground states
for some time \cite{davidnode}, namely,
that the ground state node divides the configuration space into
the minimal number of two nodal cells: a "plus" cell with $\Psi>0$ and
 a "minus" cell with $\Psi<0$.  From the node equation 
$\cos\beta=0$  of the $^3S(1s2s)$ state we easily find
that the "plus" and "minus" domains are given by the 
conditions $r_1>r_2$ and $r_1<r_2$, respectively.
For the node given by $\cos\omega'=0$ the nodal domains are given
by the orientation of ${\bf z}_0, {\bf r}_{1},{\bf r}_{2}$: the three 
vectors are either left- or right-handed, corresponding to either "plus" or "minus"
cell, respectively. Similarly, for the node $\cos\omega=0$
the domains are given by the left- or right-handedness of
the vectors ${\bf r}_{1},{\bf r}_{2}, {\bf r}_{3}$.
The minimal number of
two nodal domains
was found  
also for $2D$ and $3D$ noninteracting spin-polarized homogeneous
electron gas with periodic boundary conditions for up to 200 particles using
a numerical proof \cite{davidnode}.

Understanding the fermion nodes and their properties has become a 
challenge which might help to advance both practical calculations
but also open a deeper insights into the properties of fermionic systems.
One can envision two key problems:

a) {\it Topology of the fermion nodes, ie, how many nodal cells
are actually present} since this is of high importance for the fixed-node 
QMC approaches. Note that if the number 
of nodal cells is incorrect (typically {\it higher} than it should be since 
mean-fields such as HF 
have tendency to divide the configurations space into too many domains) then the QMC sampling
around the artificial nodes will be very sparse. This could lead to 
 large statistical fluctuations from poor sampling, 
and, possibly, to an 
effective non-ergodicity due to the finite-time projection time in practical calculations.

b) Once the topology is correct, {\it the accuracy of the manifold 
shape becomes important.} This is an area
where our insights are particularly limited since the exact nodes for 
large interacting systems are virtually unknown except
for a few-particle special cases mentioned above. 

In our recent paper \cite{mitasshort},
we have made some encouraging steps forward in trying to understand the 
topological issues
and we have 
analytically derived a number of new results regarding the number of nodal domains.
In particular, we have shown that spin-polarized 
closed-shell ground states of noninteracting harmonic 
fermions in $d=2$ and higher 
have the minimal number of two nodal domains.
We have proved the same for
spin-polarized atomic states with several electrons,
both for noninteracting and HF wave functions.
We have also shown that
by imposing additional symmetries one can generate more than two
 nodal domains but that
interactions  can relax this "nodal degeneracy" to the minimal number of two 
domains such as in the case of the $^4S(1s2s3s)$ atomic state.

For noninteracting spin-unpolarized systems, ie, with both spin channels
occupied, the number of nodal cells is four since the wave function is
a product of spin-up and spin-down Slater determinants ($2\times 2=4$).
(Here and later on we assume that the Hamiltonian does not include spin-dependent
terms so that the particle spins are conserved. 
The wave function is then a product of two determinants which depend 
only on the spatial degrees of freedom \cite{qmcrev}.)
In the last few years, studies of a few-particle systems have revealed
that interactions and many-body effects
do affect the nodal topologies
and can change the number of nodal cells \cite{dariobe,cmt28}.
In particular, for the case of Be atom it has been found that
the noninteracting/HF four nodal cells of the singlet ground state
change to the minimal number of two due to the electron correlation
\cite{dariobe} and qualitatively the same has been observed
in  other systems \cite{cmt28}. 
In our recent paper \cite{mitasshort} we have found that this is a rather
generic property of ground states in interacting systems.
 With some conditions, 
 we have explicitly demonstrated that interactions lift
the "nodal cell degeneracy" in spin-unpolarized systems
and smooth out the four noninteracting cells
into the minimal two. We have shown this for
 $2D$  harmonic fermions in closed-shell states {\it of arbitrary size}
using a variational Bardeen-Cooper-Schrieffer wave function.

In this work we further advance these ideas.
We analyze the fermion nodes in a number of other paradigmatic fermionic
models: homogeneous electron gas with periodic
boundary conditions, particles in an infinite well and on the surface of a sphere.
For all these systems
we prove that in the spin-polarized noninteracting closed-shell ground state the fermion nodes 
divide the configuration space into minimal two cells for arbitrary number of
particles.
We also extend our previous proof for spin-unpolarized systems
demonstrating 
that interactions smooth out multiple nodal cells of noninteracting/mean-field
wave functions into the minimal two for more systems such as homogeneous
electron gas and $3D$ harmonic oscillator. 
These results
 contribute to our understanding
of the fermion nodes and their impacts on the accuracy 
of wave functions with direct implications for the QMC methods. 

In the last sections we show how in periodic
spin-unpolarized interacting fermion gas the pairs of particles 
can wind around the box without crossing the node
what points towards the connection of nodes 
to the transport and to the 
existence of other quantum phases such as superconductivity.
 Finally, we then
generalize the results to the temperature density matrices 
with the implications
for path integral Monte Carlo methods \cite{davidhe}.

\section{II. General properties of fermion nodes.}

Let us introduce the basic properties of fermion nodes
as they were studied
by Ceperley \cite{davidnode} some time ago. 

a) Nondegenerate ground states wave functions fulfill the so-called 
tiling property.  Let us define a nodal cell
 $\Omega(R_0)$ as a subset of configurations which can be reached
from the point $R_0$ by a continuous path without crossing the node.
The tiling property
says that by applying all possible particle permutations to an arbitrary
nodal cell
one covers the complete configuration
space. Note that this {\em does not} specify how many nodal cells are there.

b) If $m$ nodal surfaces cross then the angle of crossing is
$\pi/m$.
 Furthermore, the symmetry of the node is the same as the
symmetry of the state.

c) 
It is possible to show that there are only two nodal cells using the following
argument based on triple exchanges. Let us first 
introduce the notion of particles connected by triple exchanges. 
We will call the three particles $i,j,k$  {\it connected}
if there exists a triple exchange path $i\to j, j\to k, k\to i$ which does 
not cross the node. If more than three particles are connected then  
they can form a connected cluster. An example of 
six particles connected into a single cluster 
is sketched as follows: 
\begin{picture}(12,7.5)(6,4.5)
\put(2,5){$\bullet$}
\put(6,2){$\bullet$}
\put(6,8){$\bullet$}
\put(10,5){$\bullet$}
\put(14,2){$\bullet$}
\put(14,8){$\bullet$}
\drawline(3.5,7)(7.7,4)(7.7,10)(3.5,7)
\drawline(7.7,10)(15.8,4)
\drawline(7.7,4)(15.8,10)(15.8,4)
\end{picture}.
If there exists a point $R_t$ such that {\it all} the 
particles are connected  into a single cluster then $\Psi(R)$ has only two nodal cells.
This can be better understood once we realize the following two facts.
First, any triple 
permutation can be written as two pair permutations. Therefore the connected
cluster of triple permutations enables to realize any even parity permutation 
without crossing the node. That exhausts all permutations which are available
for cell of one sign since the  wave function is invariant to even parity permutations. 
Second, the tiling property implies that once the particles are connected for $R_t$
the same is true for the entire cell. By symmetry, the same arguments
apply 
to the complementary "minus" cell which correspond to the odd permutations. 
More details on this property can be found in the original Ref. \cite{davidnode}.


\section{\label{sec:sp} III. Noninteracting spin-polarized fermions.}

\subsection{III.a. Homogeneous electron gas. }


We consider a
system of spin-polarized noninteracting
fermions in a periodic box  in $d$ dimensions. The  
spatial coordinates are rescaled by the box size
so that we can use dimensionless variables and the box
becomes
$(-\pi,\pi)^d$. 

We first analyze the fermion nodes for $d=1$ since the result will be
useful in subsequent derivations. We consider 
a system with $N=(2k_F+1)$ particles. In our $1D$ dimensionless
units the Fermi momentum becomes an integer,
$k_F=1,2 ...$.  The one-particle occupied states are written 
as $\phi_n(x)=e^{inx},$ $n=0,\pm 1,...,\pm k_F$ 
and the spin-polarized ground state is given by
\begin{equation}
\Psi_{1D}(1,...,N)={\rm det}\left[ \phi_n(x_j)\right]
\end{equation}
 where
$x_j$ is the $j$-th particle coordinate and $j=1,...,N$.
We factorize the term 
$\exp(-ik_F\sum_jx_j)$ out of the determinant so that the 
Slater matrix elements become powers of $z_j=e^{ix_j}$. 
The resulting Vandermonde determinant can be 
written in a closed form and after some rearrangements we find 
\begin{equation}
\Psi_{1D}(1,...,N)=
e^{-ik_F\sum_jx_j}\prod_{j>k}(z_j-z_k)=\nonumber
\end{equation}
\begin{equation}
=\mu_0\prod_{j>k}
\sin(x_{jk}/2)
\label{eq:van1d}
\end{equation}
where 
$x_{jk}=x_j-x_k$ and $\mu_0$ is
a constant prefactor which is unimportant 
for our purposes.  (In the derivations below we will be using
the letter $\mu$ for denoting prefactors which are either
constants or nonnegative functions in particle coordinates
and therefore do not affect the nodes).  
The derived wave function has the following important properties.

a) The fermion nodes
appear at the particle coincidence points. This implies a well-known 
result that the ground state wave function in $1D$ have $N!$ nodal cells since any 
particle permutation  
requires crossing the node at least once. In addition, this also means that 
{\it any} fermion configuration which 
preserves the particle order is contained within the same nodal domain/cell.


b) The wave function is invariant to
translations and cyclic exchanges of particles. 
Due to the periodic boundary conditions
this includes also winding the system
around the periodic box, ie, the translation of all particles
by $2\pi$. We can formalize 
this by introducing an operator $T^{x}_a$ which translates all the particles 
along the $x$-axis
as $x_j \to x_j+a$ so that the translation invariance can be written as
\begin{equation}
T_a^x \Psi_{1D}(1,...,N)= \Psi_{1D}(1,...,N)
\label{eq:inv1}
\end{equation}

Similarly, the wave function  is invariant to 
cyclic exchange of all the particles given by
$j \to j+1, j=1, ..., N$ and $N+1 \to 1$.
 This action is carried out   
by $C_{+1}^{x}$ operator where the notation means the 
exchange by one site is in the $+x$-direction. 
Clearly, the inverse operator
$[C_{+1}^x]^{-1}=C_{-1}^x$. Note that here we have assumed that $N$ is odd, in agreement with our 
definition.  The invariance holds only for $N$ odd since then
the cyclic exchange is equivalent
to an even number of pair exchanges. For the sake of completeness, we consider
also $N$ even, when the cyclic exchange
can be replaced by an odd number of pair exchanges, resulting
in the wave function sign flip.
In general, for the cyclic exchanges we can therefore write  
\begin{equation}
C_{\pm 1}^x\Psi_{1D}(1,...,N)=(-1)^{N+1}\Psi_{1D}(1,...,N)
\label{eq:inv2}
\end{equation}

c) Assuming the particle positions are all distinct, 
the cyclic exchange {\it path} can be chosen in such 
 a way that it
{\it does not cross the node}. 
This is easy to accomplish by maintaining 
finite distances between the particles along the path trajectory. 
Let us parametrize the exchange path by a parameter $t$ so that the
path starts at
 $t=0$, the path is completed at $t=1$ and the  
 exchange path operator is then denoted as $C^x_{+1}(t)$ with $0\leq t\leq 1$.
The fact that the path  
does not cross the node (ie, the path is contained within the same 
nodal cell) can be then written as
\begin{equation}
|C_{+1}^x(t)\Psi_{1D}(1,...,N)|>0, \; 0\leq t \leq 1
\label{eq:inv3}
\end{equation}
where, of course, $N$ is assumed to be odd.

Let us now derive the wave functions and generalize the results for 
fermions in $2D$. The one-particle states in $2D$ are  
$\phi_{nm}(x,y)=e^{i(nx+my)}$.  
The states are occupied up to the Fermi momentum $k_F$ so that
we have $n^2+m^2\leq k_F^2$, where $k_F$ in $2D$ is not necessarily an integer. 
Similarly to our previous paper \cite{mitasshort},
we show that the spin-polarized electron gas  
for closed-shell
ground
states have only two nodal cells.

\begin{figure}[ht]
\centering
\begin{tabular}{ccc} 
\includegraphics[width=1.1in,clip]{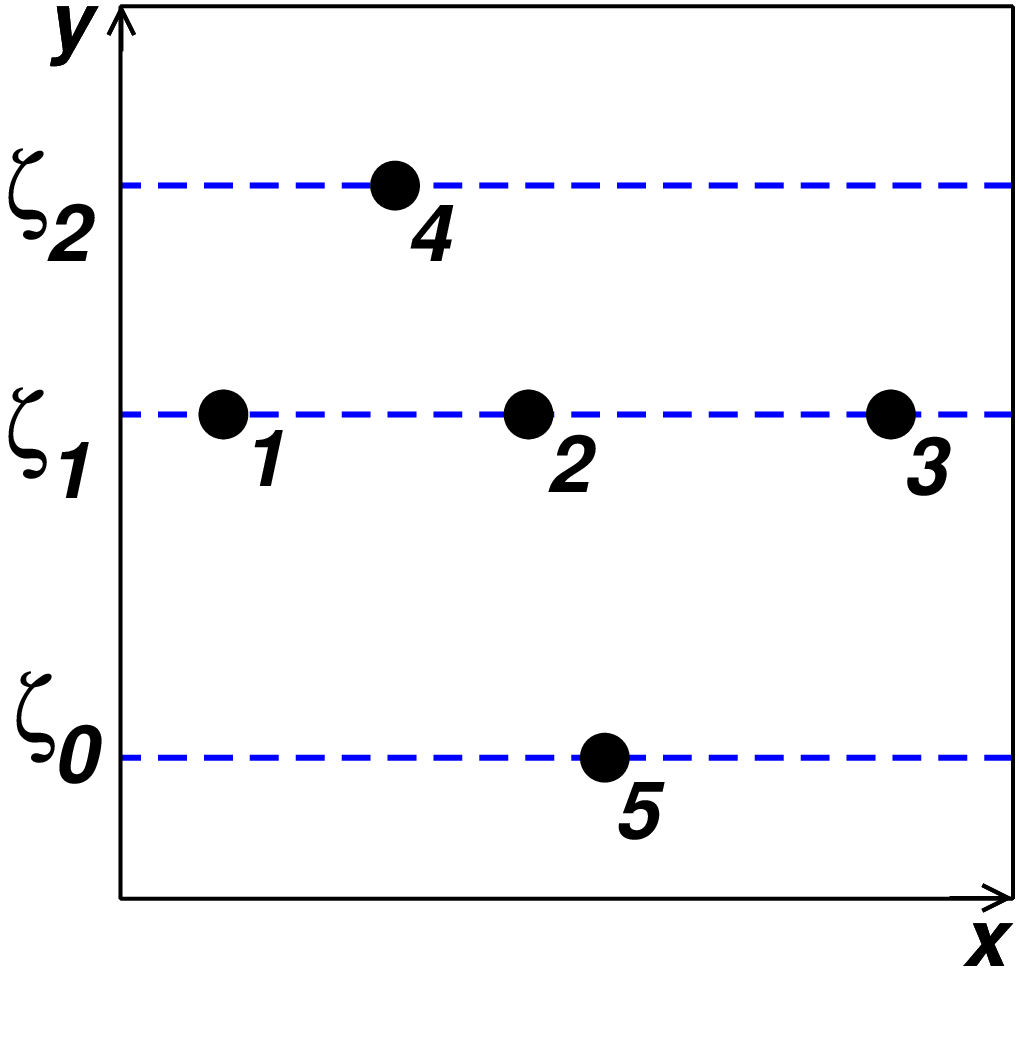} &
\includegraphics[width=1.1in,clip]{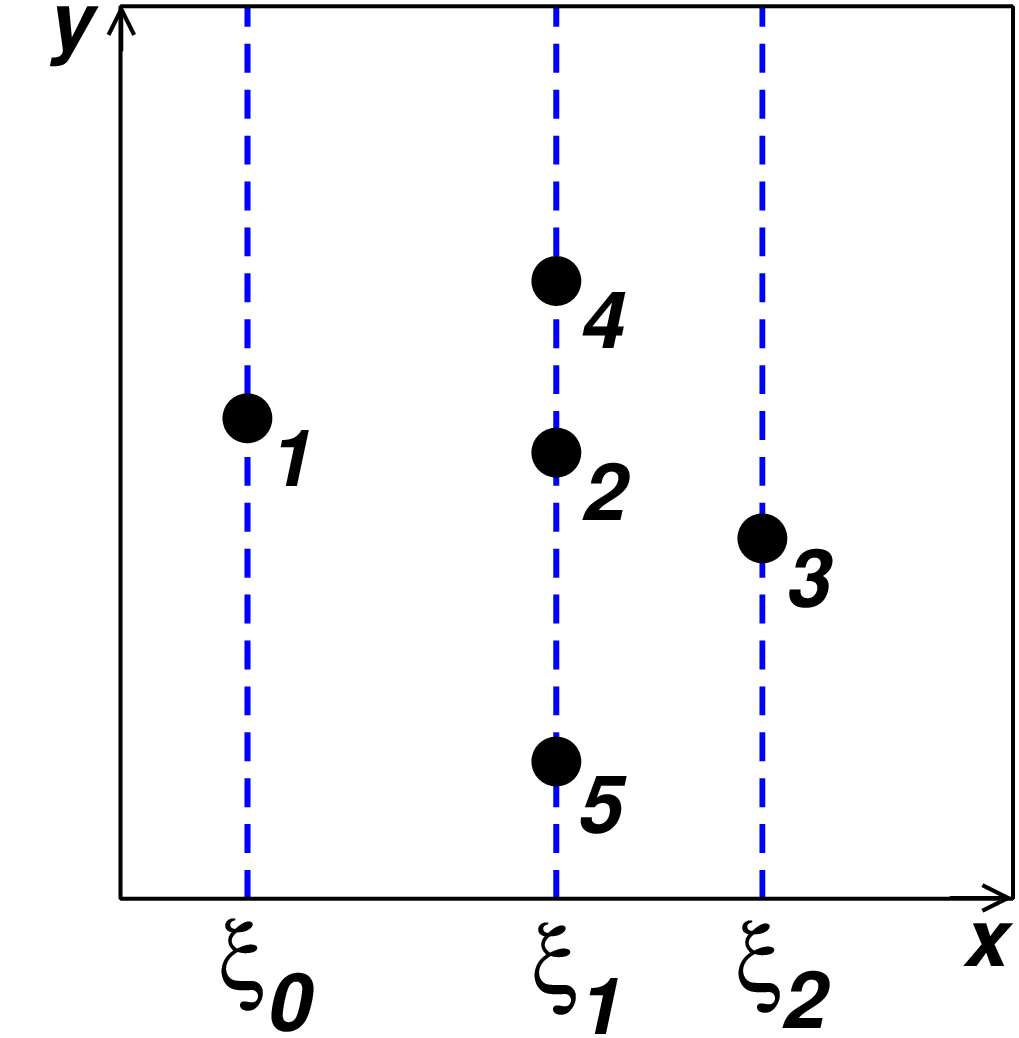} &
\includegraphics[width=1.1in,clip]{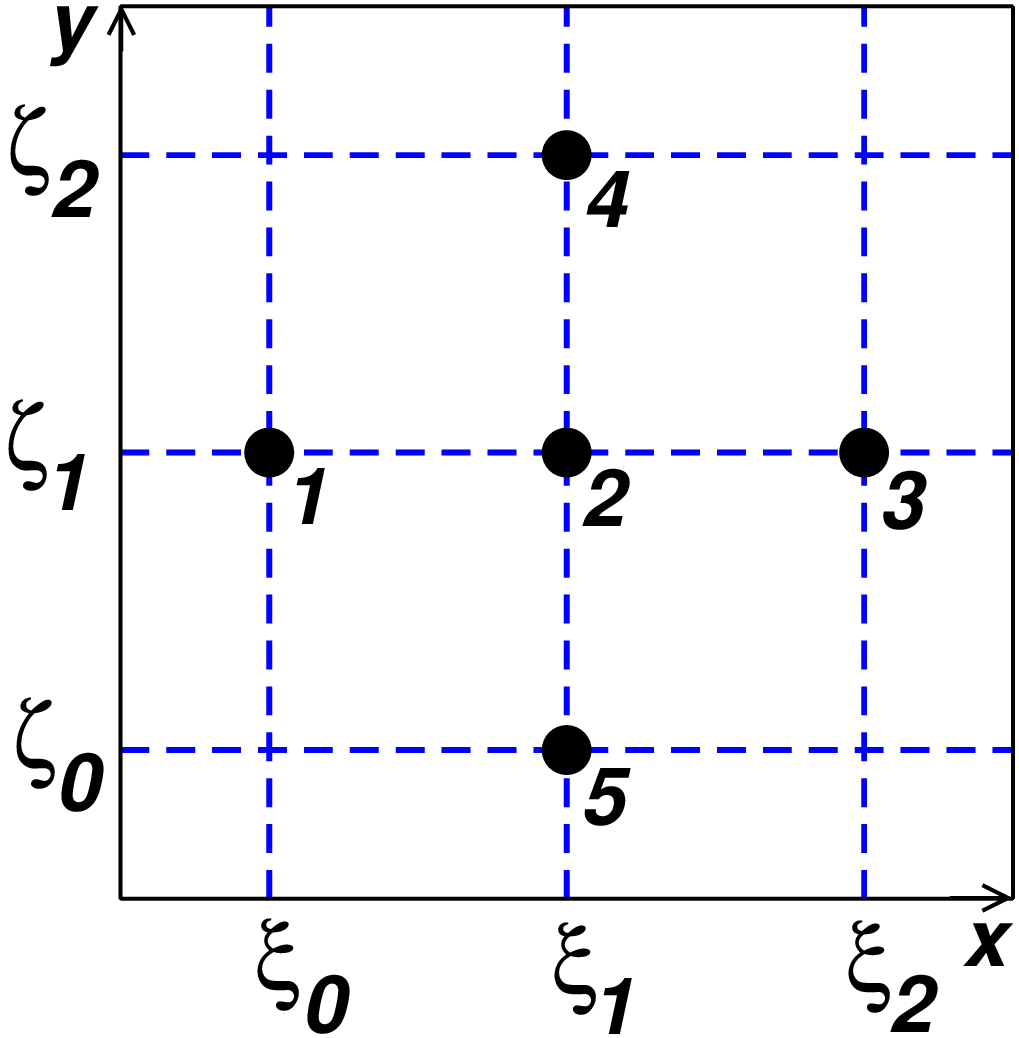} \\
(a) & (b) & (c)
\end{tabular}
\caption{Positions of five fermions in the $2D$
 periodic box $(-\pi,\pi)^2$. (a) Particles aligned horizontally,
along the lines $y=\zeta_0,\zeta_1,\zeta_2$.
(b) Particles aligned vertically, along the lines  $x=\xi_0,\xi_1,\xi_2$.  (c) Particles aligned 
in both directions, positioned on a lattice.}
\label{fig:five2d}
\end{figure}

The proof is by induction, therefore 
let us first consider $k_F=1$ with five particles
occupying $\{1,e^{\pm ix},
e^{\pm iy}\}$ one-particle states. 
We place the particles as in Fig.\ref{fig:five2d}a so that
the coordinates are given by 
${\bf r}_1=(x_1,\zeta_1)$, ${\bf r}_2=(x_2,\zeta_1)$, ${\bf r}_3=(x_3,\zeta_1)$,
 ${\bf r}_4=(x_3,\zeta_2)$,  ${\bf r}_5=(x_5,\zeta_0)$.
By eliminating the terms with $\zeta_1$ 
the wave function can be factorized 
as follows  
\begin{equation}
\Psi_{2D}(1,...,5)=\left| \begin{array}{ccccc}
   1  &  1   &  1   &    1  &  1    \\
 e^{ix_1}  &  e^{ix_2}   &  e^{ix_3}   &
   e^{ix_4}  &  e^{ix_5} \\
 e^{-ix_1}  &  e^{-ix_2}   &  e^{-ix_3}   &
   e^{-ix_4}  &  e^{-ix_5} \\
 e^{i\zeta_1} &  e^{i\zeta_1} &
e^{i\zeta_1} &  e^{i\zeta_0} &
e^{i\zeta_2} \\
 e^{-i\zeta_1} &  e^{-i\zeta_1} &
e^{-i\zeta_1} &  e^{-i\zeta_0} &
e^{-i\zeta_2} \\
\end{array} \right|=\nonumber
\end{equation}
\begin{equation}
\mu_0\left| \begin{array}{ccc}
   1  &  1   &  1     \\
 e^{ix_1}  &  e^{ix_2}   &  e^{ix_3}  \\
 e^{-ix_1}  &  e^{-ix_2}   &  e^{-ix_3}  \\
\end{array} \right|
\left| \begin{array}{cc}
  e^{i\zeta_0}-e^{i\zeta_1} &
e^{i\zeta_2} -e^{i\zeta_1}\\
 e^{-i\zeta_0} -e^{-i\zeta_1} &
e^{-i\zeta_2} -e^{-i\zeta_1} \\
\end{array} \right|=\nonumber
\end{equation}

\begin{equation}
= \mu_0'\Psi_{1D}(1,2,3)\sin(\zeta_{10}/2)\sin(\zeta_{20}/2)\sin(\zeta_{21}/2)
\label{eq:five2dzeta}
\end{equation}
where $\mu_0,\mu_0'$ are irrelevant constant prefactors.
We have obtained a product of $1D$ wave function and
terms with distances between the
{\it lines} $y=\zeta_0,\zeta_1,\zeta_2$.
Note that analogous result can be found for the configuration
sketched in Fig. 1b so the particles are aligned 
in parallel to the 
$y$-axis with the wave function given by
\begin{equation}
\Psi_{2D}(1,...,5)=
 \mu_0\Psi_{1D}(5,2,4)\times \nonumber
\end{equation}
\begin{equation}
\times \sin(\xi_{10}/2)\sin(\xi_{20}/2)\sin(\xi_{21}/2)
\label{eq:five2dxi}
\end{equation} 
If the particles are aligned in both directions as in Fig. 1c
both expressions apply.
From the derived wave function it is clear that the node is encountered
when particles lying on the same line reorder or 
when the lines of particles 
cross each other, eg, $\xi_1=\xi_0$.
The wave functions for the configurations outlined 
in Fig.\ref{fig:five2d} possess an  
important property. Consider 
$\Psi_{1D}(...)$ in Eqs.  \ref{eq:five2dzeta},  \ref{eq:five2dxi} with
 groups of particles positioned on the corresponding lines.
Assuming the group of particles is allowed to move only along the given line,
one can consider this to be an effective $1D$ subspace.
The $1D$ wave function for such a subspace  
obeys all three conditions derived
for $1D$ case, ie, 
 Eqs. \ref{eq:inv1}, \ref{eq:inv2} and \ref{eq:inv3}. 
For example, for $\Psi_{1D}(1,2,3)$ which appears 
 in Eq. \ref{eq:five2dzeta} we have
 $|C_{+1}^x(t)\Psi_{1D}(1,2,3)|>0$. This is easy to check also analytically. 
We fix the coordinates in 
 Fig.1(a) as follows $x_1=-2\pi/3, x_2=0, x_3=2\pi/3$ 
and then we can 
carry out a "synchronized" cyclic exchange using the translation  
by $2\pi/3$ in $x$-direction. The 
wave function is constant 
along the translation/cyclic exchange path so that we have 
\begin{equation}
C^x_{+1}(t) \Psi_{1D}(1,2,3)=\Psi_{1D}(1,2,3), \;\; 0\leq t \leq 1
\end{equation}
because of the translational invariance.

Finally, we are ready to show that there are only two nodal cells in this 
five-particle ground state. 
Assume that the particles are positioned as in Fig.1c. 
Consider the following sequence of 
four exchanges, $C^x_{+1}C^y_{+1}C^x_{-1}C^y_{-1}$, where the operators act
from the right, ie, $C^y_{-1}$, acts first.  We denote this exchange symbolically as 
\begin{equation}
C_{\rightarrow\uparrow\leftarrow\downarrow}=C^x_{+1}C^y_{+1}C^x_{-1}C^y_{-1}
\end{equation}
The exchanges are performed only for  
the particles lying on the lines $\zeta_1,\xi_1$ in corresponding directions,
eg, $C_{\pm 1}^x$ acts on particles along  $y=\zeta_1$ while  $C_{\pm 1}^y$ 
cycles particles along $x=\xi_1$ . 
It is easy to check that this results in a triple exchange 
$423 \to 342$ while the particles 1 and 5 end up in their original
positions. It is therefore clear that particles 2,3,4 are connected by 
this triple exchange.
It is straightforward to show that 
$C^x_{-1} C^y_{-1} C^x_{+1}C^y_{+1}$ exchanges particles 1,2,5 so that 
we can conclude that all five particles are connected by triple exchanges    
into a single cluster
and there are only two nodal domains.

Now we need to extend the arguments and find the wave functions for
general cases 
with occupied states within the Fermi disk with an arbitrary $k_F$.
It is instructive to first derive 
the wave functions for $k_F=\sqrt{2}$ with 9 particles
occupying
 states $\{ 1,e^{\pm ix}, e^{\pm iy}, e^{i(\pm x \pm y)}\}$
 and 
for $k_F=2$ with 13 particles  with additional states 
$\{ e^{\pm i2x}, e^{\pm i2y}\}$.
This will enable us to understand how to  
write down the wave function
for a general case while avoiding rather tedious notations 
which would appear in the direct derivations.

\begin{figure}[ht]
\centering
\begin{tabular}{ccc}
\includegraphics[width=1.35in,clip]{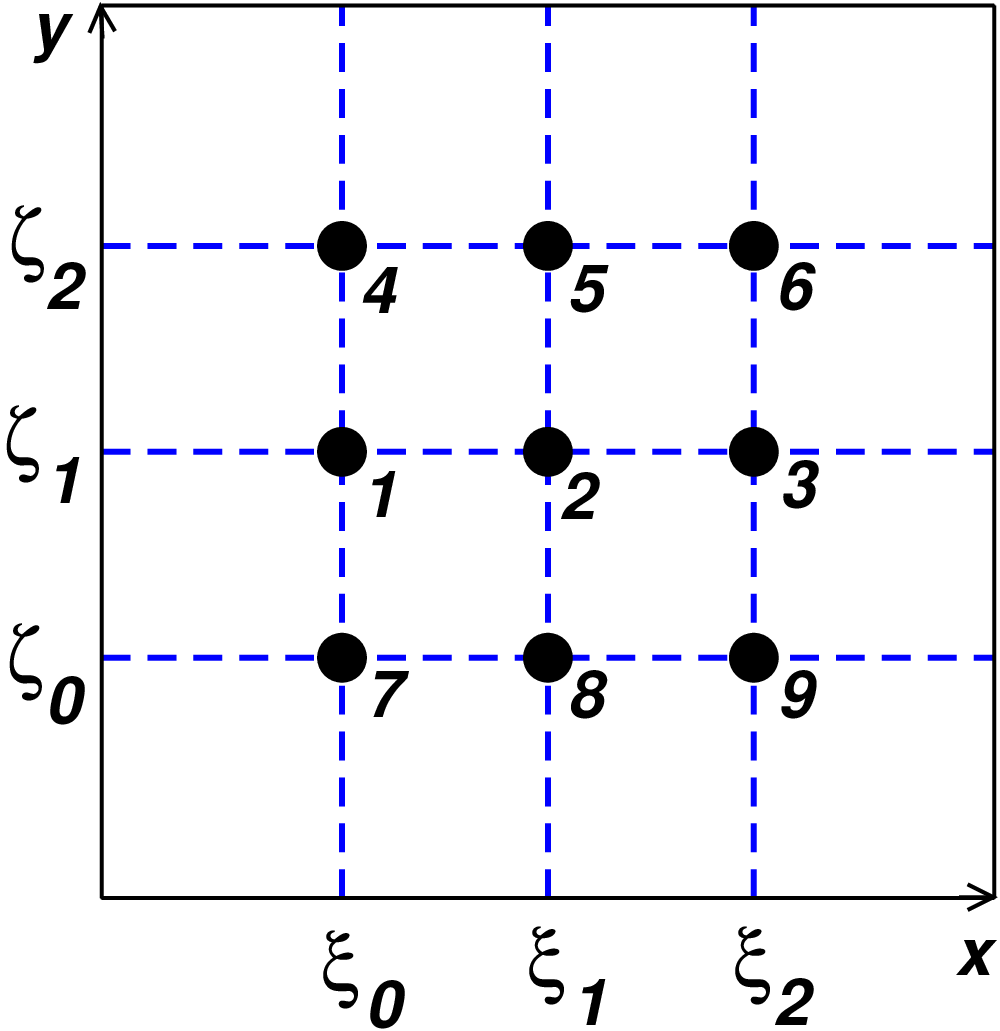} &
\quad &
\includegraphics[width=1.35in,clip]{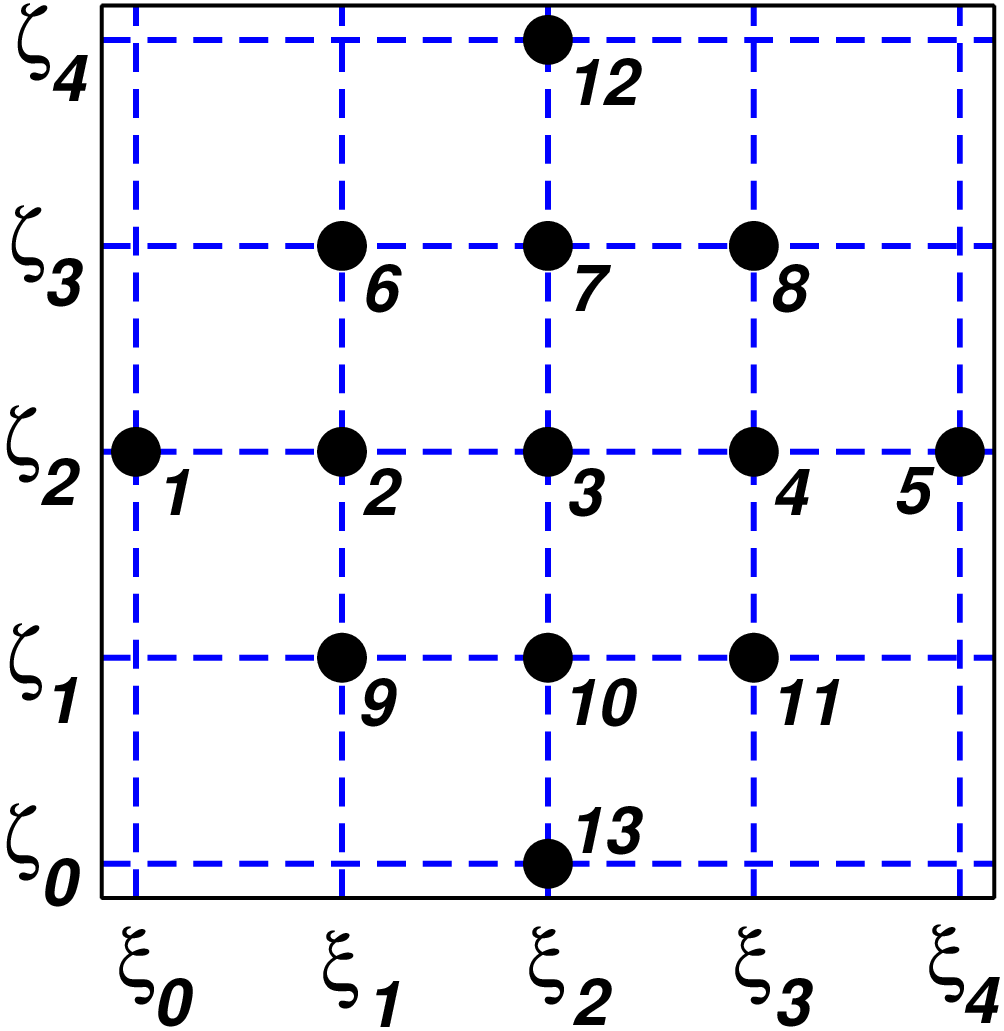} \\
(a) & \quad & (b)
\end{tabular}
\caption{Positions of fermions in
 the $2D$ periodic box in real space. (a) Positions of the nine-fermion system.
(b) Positions of the 13-fermion system.
See the text for details. }
\label{fig:nine2d}
\end{figure}

 Let us position the particles into a pattern which mimics the lattice
of the occupied wave vectors in the reciprocal space, see Fig.\ref{fig:nine2d}.
Using appropriate algebraic arrangements we subsequently factorize the lines 
of particles as given by
\begin{equation}
\Psi_{2D}(1,...,9)= \nonumber
\end{equation}
\begin{equation}
=\mu_0\Psi_{1D}(1,2,3)\Psi_{2D}(4,...,9) \sin^3(\zeta_{21}/2) \sin^3(\zeta_{10}/2)=\nonumber
\end{equation}
\begin{equation}
=
\mu_0'\Psi_{1D}(1,2,3)\Psi_{1D}(4,5,6)\Psi_{1D}(7,8,9)\times \nonumber
\end{equation}
\begin{equation}
\times \sin^3(\zeta_{21}/2) \sin^3(\zeta_{10}/2)\sin^3(\zeta_{20}/2)
\end{equation}
where $\mu_0,\mu_0'$ are constants.
Similarly, for $k_F=2$ we can first factorize
the line with the largest number of particles on 
$y=\zeta_2$ (see Fig.\ref{fig:nine2d})
\begin{equation}
\Psi_{2D}(1,...,13)=
\mu_0\Psi_{1D}(1,...,5)\sin(\zeta_{02}/2)\sin^3(\zeta_{12}/2)
\times
\nonumber
\end{equation}
\begin{equation}
\times \sin^3(\zeta_{32}/2)\sin(\zeta_{42}/2)
\Psi_{2D}(6,...,13)\nonumber
\end{equation}
and 
then factorize out the particles lying on another line $y=\zeta_3$ to obtain
\begin{equation}
\Psi_{2D}(1,...,13)=
\mu_0\left[\prod_{j\neq 2}\sin^{n_j}(\zeta_{j2}/2)\right]
\left[\prod_{j\neq 2,3}\sin^{n_j}(\zeta_{j1}/2)\right]
\times
\nonumber
\end{equation}
\begin{equation}
\Psi_{1D}(1,...,5)\Psi_{1D}(6,7,8)\Psi_{2D}(9,...,13)
\end{equation}
where $n_j$ is the number of particles lying on the line $y=\zeta_j$.
We have obtained a product of $1D$ wave functions, 
terms with distances 
between the lines and the $2D$ wave function with lower number of particles
positioned in the same type of pattern is in Fig.\ref{fig:five2d}. Obviously, we can
further factorize  $\Psi_{2D}(9,...,13)$ using Eq.\ref{eq:five2dzeta} until
we end up 
with $1D$ factors only.
Using these insights into the recursive form of the wave function, 
for a general case with $M+1$ lines we can write
\begin{equation}
\Psi_{2D}(1,...,N)=\nonumber
\end{equation}
\begin{equation}
=\mu_0\prod_{k=0}^{M}\left[\Psi_{1D}(I_k)\prod_{j>k } 
\sin^{n_j}(\zeta_{jk}/2)\right]
\end{equation}
where $I_k=i^{(k)}_1, ..., i^{(k)}_{n_k}$
denotes the labels of particles
lying on the line $y=\zeta_k$.
In addition, analogous expression can be found if we factorize along the $x=\xi_j$
lines, the only difference being replacement $\zeta_{jk}$ by $\xi_{jk}$ 
and corresponding replacement of particle sets in $\Psi_{1D}$ wave functions.

\begin{figure}[ht]
\centering
\begin{tabular}{ccc}
\includegraphics[width=1.35in,clip]{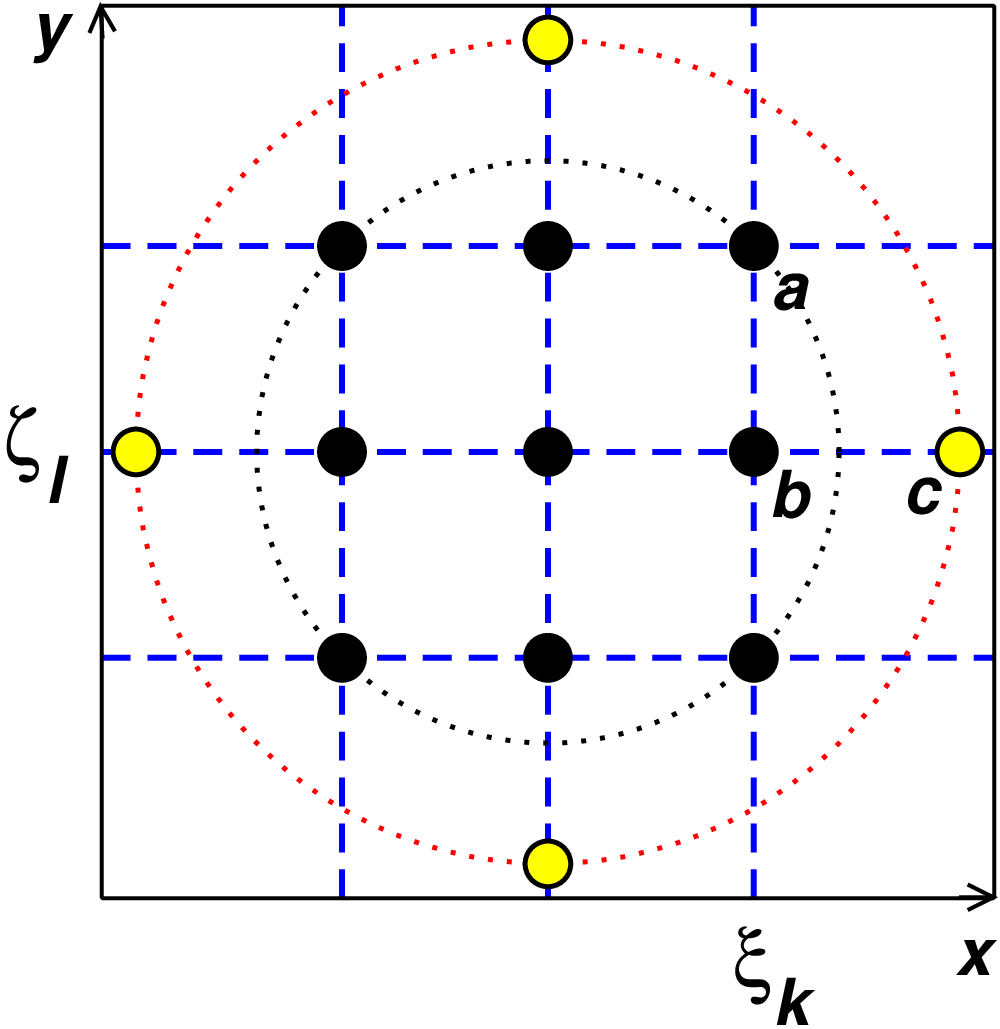} &
\quad &
\includegraphics[width=1.35in,clip]{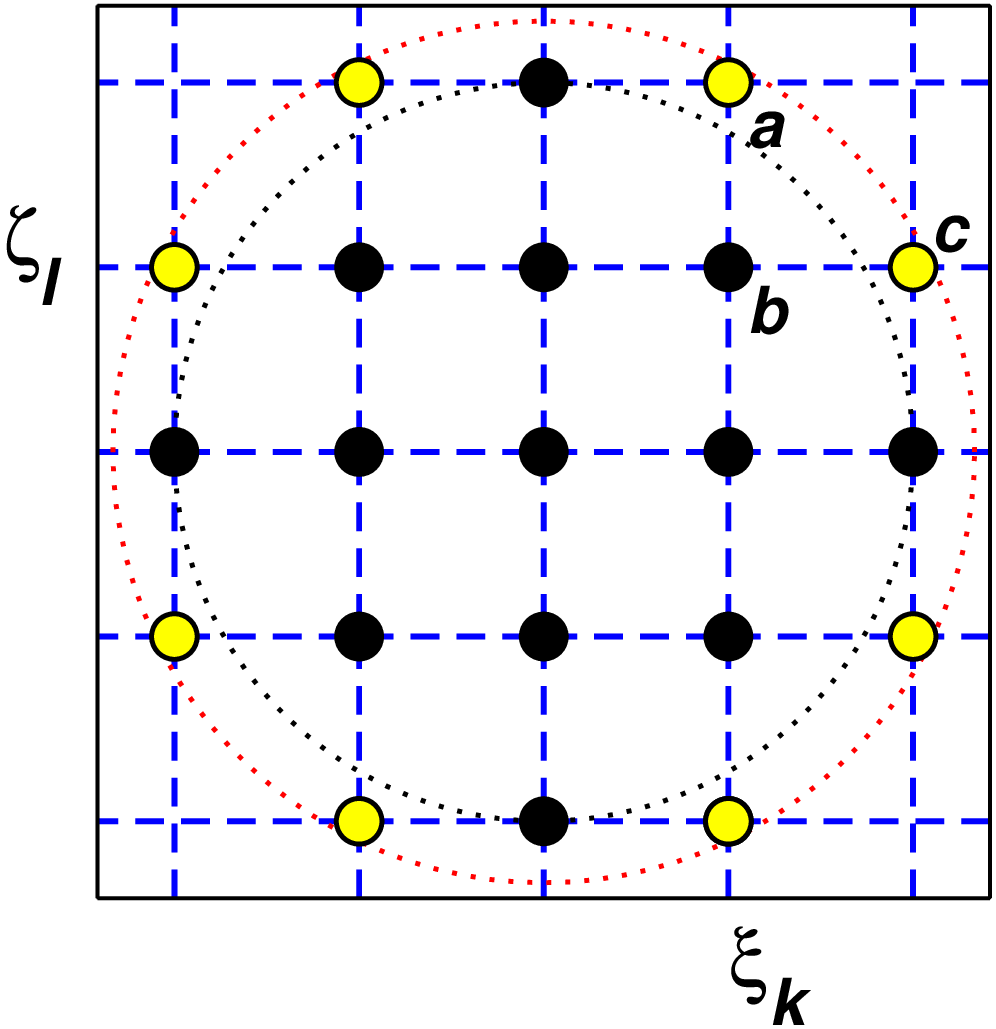} \\
(a) & \quad & (b)
\end{tabular}
\caption{Illustrations of the position patterns in $2D$ periodic box for the
 size increases from $k_F$ to $k_F+\Delta k_F$.
The particle layouts mimic the occupied states in the
reciprocal space, however, the particles are positioned in the real space. 
The additional particles are in grey/yellow.
(a) The additional particles positioned along $x-$ and $y-$axis directions.
(b)  The additional particles positioned along diagonals.
Lines $x=\xi_k$ and $y=\zeta_l$ are used for the proof that the particles $a,b,c$
are connected by a triple exchange.
See the text for details. }
\label{fig:indu2d}
\end{figure}

We are now ready for the induction step. Consider a 
spin-polarized closed-shell ground state
with a given $k_F$. For this wave function we assume that all the
particles are connected by the triple exchanges, ie, there are only two nodal domains.
Let us
increase $k_F \to k_F+\Delta k_F $ until the Fermi disk
includes the next unoccupied star of states $\phi_{nm}(x,y)$ for which
$k_F < (n^2+m^2)^{1/2} \leq k_F+\Delta k_F$. The system size increases
by the corresponding number of additional particles.
Assuming the particles are positioned in the real space 
in the same pattern as
 the occupied ${\bf k}$-points in reciprocal space (Fig.\ref{fig:indu2d}),
 the additional particles will appear at the 
borderline of the disk in the real space.

The two basic possibilities how the additional particles are positioned 
are  given in the Fig.\ref{fig:indu2d}.  
We need to show that these additional particles are connected 
to the original particles
by the triple exchanges. 
 This can be demonstrated by
the sequence of the four cyclic exchanges which we have
used above for the five particle case. 
It involves particles on
  the lines $x=\xi_k$ and $y=\zeta_l$ as schematically drawn in Fig.\ref{fig:indu2d}.
Consider the exchange
$C_{\rightarrow\uparrow\leftarrow\downarrow}=$
$C^x_{+1}C^y_{+1}C^x_{-1}C^y_{-1}$  where the cyclic exchanges in $x$ direction are
applied only to particles on the line $x=\xi_k$ and, similarly, 
the cyclic exchanges in $y$ direction are applied only to particles along $y=\zeta_l$. 
 Note the wave function is invariant to cyclic exchanges
since the number of particles along each line 
is odd for any closed shell state. It is then straightforward to find 
out that $C_{\rightarrow\uparrow\leftarrow\downarrow}$
 exchanges particles $a,b,c$ while the rest of the particles remains intact.
 Similar exchanges can be carried out for 
all additional particles. Finally, this shows
that the additional particles are connected 
to
the particles of the wave function with size $k_F$ and finalizes the 
proof.

The proof can be extended into $3D$ and higher dimension by positioning the particles onto
an appropriate $3D$ pattern which in real space mimics the occupied states in the Fermi
sphere in the reciprocal space. This is possible due to the fact that with proper positioning
of particles
one can subsequently factorize the Slater
determinant along {\it hyperplanes, planes and lines}. Using arguments similar to the $2D$ case 
we can perfrom cyclic exchanges without
crossing the node in $3D$ and higher dimensions. Therefore the proof for higher dimensions
is essentially the same. 

In many cases,
the proof can be extended to open shells. If an open-shell state is degenerate
it is necessary to fix 
the ambiguity in the nodes, for example, by considering
wave functions which transform according to an appropriate
symmetry subgroup \cite{davidnode, foulkes}. In general, however, depending on symmetries,
the number of degenerate
states, etc, one cannot rule out possibilities  
of states with the number of
cells beyond the minimal two  (this will be investigated in the next paper).
 
This concludes the arguments and the proof that for $d>1$ the 
noninteracting spin-polarized
fermion gas closed-shell ground states in periodic boundary conditions have only two nodal cells. 

\smallskip
\subsection{ III.a. Fermions on the surface of a sphere.}

\begin{figure}[ht]
\centering
\begin{tabular}{ccc}
\includegraphics[width=1.35in,clip]{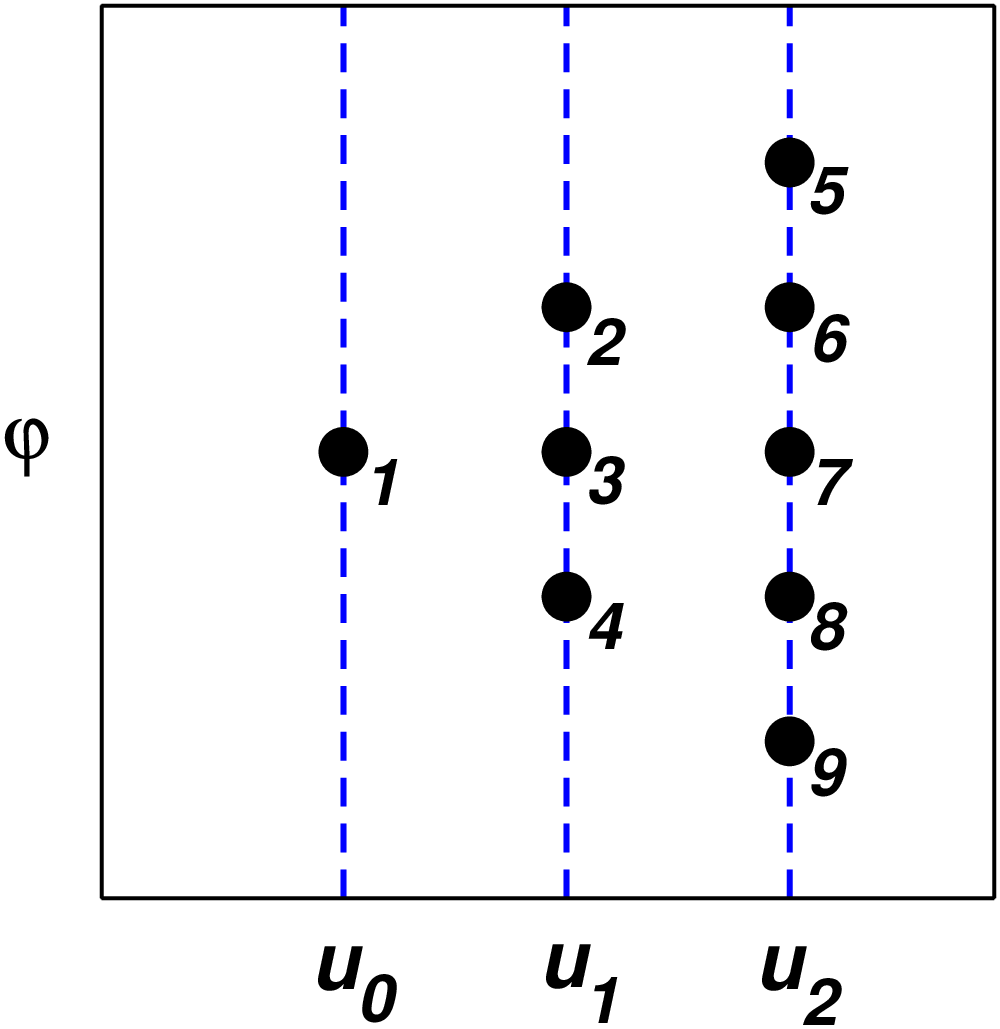} &
\quad &
\includegraphics[width=1.35in,clip]{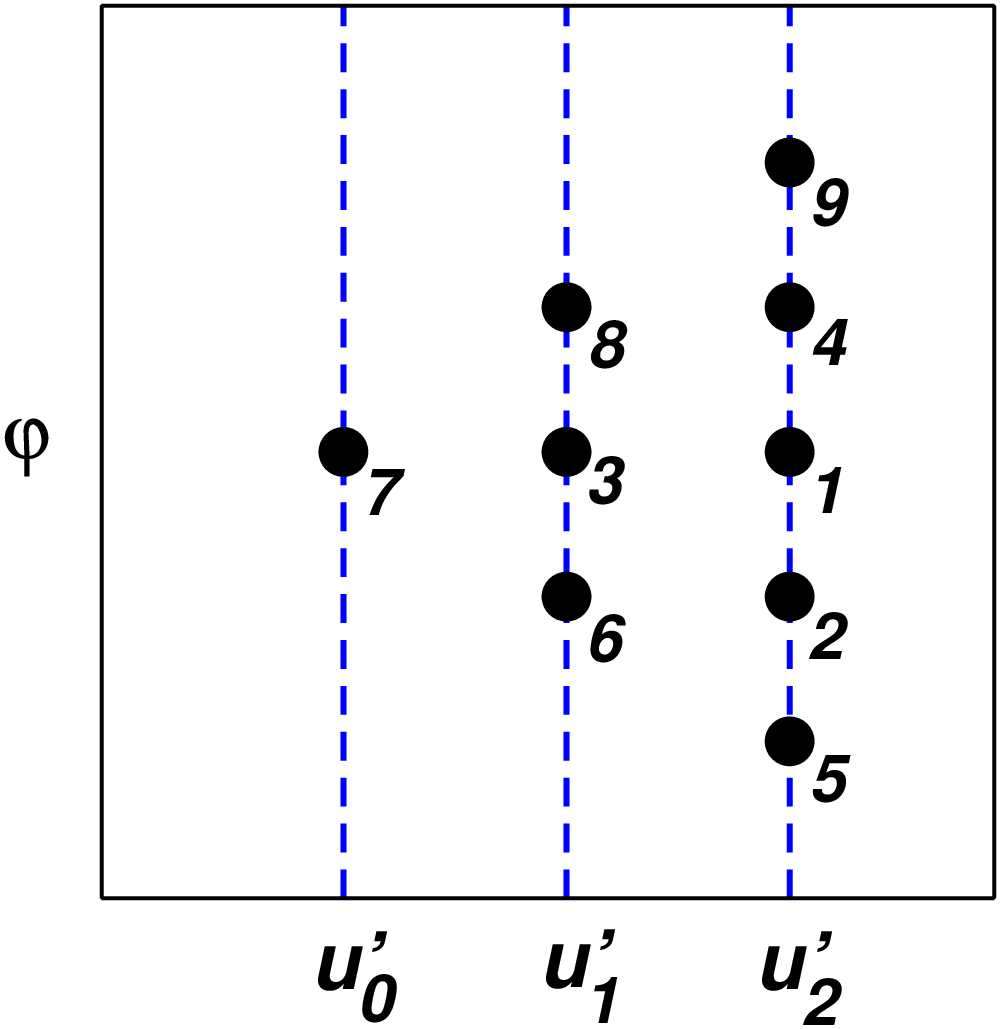} \\
(a) & \quad & (b)
\end{tabular}
\caption{
Positions of the particles on the sphere surface for the state $^{10}S(sp^3d^5)$
using coordinate system $u=\cos(\vartheta)$
and $0\leq \varphi \leq 2\pi$.
The particles are positioned along the lines with constant $u=u_j$.
For larger systems the patterns are the same with particles corresponding to increasing
angular momentum $l$ lying on the lines with increasing $u_l$. (a) Particle positions
used for derivation of the wave function factorization. (b) Particles after appropriate
shifts along $\varphi$ and a subsequent rotation.  }
\label{fig:sph}
\end{figure}

 Spin-polarized free fermions on the surface of a sphere are 
described by a Slater determinant with 
one-particle states being the spherical harmonics
$Y_{lm}(\vartheta,\varphi)$. The spherical harmonics are polynomials 
in variables $\cos\vartheta$ and $\sin\vartheta e^{\pm i\varphi}$. 
 The Slater matrix elements can be  
rearranged to monomials in these variables and it 
is then straightforward to
factorize the determinant into similar form as obtained
for the homogeneous fermion gas or for the harmonic oscillator
\cite{mitasshort}. Let us assume that the particles are positioned
as sketched in Fig.~\ref{fig:sph}a.  Using familiar expressions for
the first few spherical harmonics we get for the $^5S(sp^3)$ state 
\begin{equation}
\Psi_{sp^3}(1,...,4)=\mu_0(u_1-u_0) v_1^2\prod_{2\leq i<j\leq 4}\sin(\varphi_{ij}/2)
\end{equation}
where we have denoted $u=\cos\vartheta$ and $v=\sin\vartheta$ and $\mu_0$ 
is a constant. Similarly,
for the  $^{10}S(sp^3d^5)$ state with nine fermions we obtain
\begin{equation}
\Psi_{sp^3d^5}(1,...,9)=
 \mu'_0\Psi_{sp^3}(1,...,4) \times \nonumber
\end{equation}
\begin{equation}
\times (u_2-u_0)(u_2-u_1)^3 v_2^6 \Psi_{1D}(5,...,9)
\end{equation}
where we have denoted 
$\Psi_{1D}(1,...,N)=
\prod_{j<k}\sin(\varphi_{jk}/2)$ in agreement with Eq. \ref{eq:van1d}.

For an arbitrary size closed-shell $S$ symmetry state the one-particle 
states are occupied up to the angular
momentum $L$ with the corresponding number of particles  
$N=\sum_{l=0}^L(2l+1)=(L+1)^2$. 
 Using the previous two examples 
we can directly write the factorized wave function assuming  
the
particle positions follow the pattern in Fig.~\ref{fig:sph}a. 
We first factorize the line $u=u_L$ and then recursively the rest of the lines
so that we can write
\begin{equation}
\Psi(1,...,N)= \mu_0\Psi_{2D}(1,...,N/I_L)\times
\nonumber
\end{equation}
\begin{equation}
\times \Psi_{1D}(I_L)v_L^{L(L+1)}\prod_{0\leq j<L}(u_L-u_j)^{n_j}=\nonumber
\end{equation}

\begin{equation}
=\mu(v_1, ...,v_L)\prod_{k=0}^{L-1}\left[
\Psi_{1D}(I_{L-k}) 
\prod_{j<L-k}(u_{L-k}-u_j)^{n_j} 
\right]
\end{equation}
where $I_k$ is the subset of particles lying on the line $u=u_k$.
 In the final expression the powers of $v_j$, 
such as $v_L^{L(L+1)}$ in the preceding line, are absorbed
into $\mu(v_1, ..., v_L)$.  This prefactor is nonnegative
and does not affect the nodes or the wave function rotational invariance.

We use the properties of $1D$ wave functions and 
the rotational invariance to build the proof of the two nodal domains. First,
note that for four particles in $^5S(sp^3)$ state it is easy to show \cite{mitasshort}
that there are only 
two domains. A little bit of algebra shows that
 the node is encountered when all four particles lie on a circle resulting from a
plane cutting the sphere.  Clearly, it is easy to position the four particles 
on the sphere so that the triple exchanges do not cross this node.  For the induction
step assume that for the size $L$ with $N=(L+1)^2$ the particles are connected. We need to show
that for the size with $L \to L+1$ with additional $2(L+1)+1$ particles, the additional 
particles are connected as well. Using the fact that  particles can be shifted along
the $\varphi$ coordinate and rotated, one can reposition the particles as illustrated
on the $^{10}S(sp^3d^5)$ state,  Fig.~\ref{fig:sph}b. 
By applying the factorization to this particle arrangement we see that
the additional particles are connected and the argument applies to an arbitrary
size closed-shell state.
 

\subsection{\label{sec:boxsp} III.c. Fermions in a box.}
Let us assume a system of fermions in
a box $(0,\pi)^d$ with the condition that the wave function vanishes
at the boundaries. For $d=1$ the one-particle
states are $\phi_n(x)=\sin(nx), n=1,2,... $. Note that this can be written as 
$\phi_n(x)=\sin(x)U_{n-1}(\cos x)$ where $U_n$ is the $n$-th degree Chebyshev polynomial of 
the second kind. 
We map the variables $x_i$ to $p_i=\cos(x_i)$  so that 
$p_i\in (-1,1)$ for $i=1,...,N$. Note that the map $x \to p$ 
is a homeomorphism (ie, it is bijective and continuous with its inverse). 
Homeomorphisms
 preserve topologies, eg, 
the ordering of points, so that $x_a<x_b$ $\Leftrightarrow$ $p_a<p_b$.
Using this map we can
write the wave function for $N$ fermions in the $1D$ box directly as 
\begin{equation}
\Psi_{1D}(1,...,N)={\rm det} [\phi_n(x_i)] 
=\mu_0\prod_k \sin(x_k) \prod_{i>j}(p_i-p_j) 
\end{equation}
where $\mu_0$ is a constant.

In the $2D$ box the one-particle states are given by 
$\phi_{nm}(x,y)=\sin(x)\sin(y)U_{n-1}(p)U_{m-1}(q),
\; n,m=1,2, ...$ where we denoted $q=\cos(y)$.
If we absorb $\sin(x)\sin(y)$ into the common prefactor $\mu$,
the Slater matrix
 elements become monomials of the type $p^{n-1}q^{m-1}$. 
The states which are occupied lie within the quarter of 
the Fermi disk $n^2+m^2 \leq k_F^2$ and $ n,m>0$. 
Assume that the particles lie on a lattice in the space of variables $p,q$ 
so that they are positioned 
on $M$  horizontal and vertical lines (see, eg, the upper right quadrant in Fig.3c).
Using the techniques presented above and in the
previous paper \cite{mitasshort} we can write down the wave function as 
\begin{equation}
\Psi_{2D}(1,...,N)= 
\nonumber
\end{equation}
\begin{equation}
= \mu(x_1,y_1,...,x_N,y_N)\prod_{k=1}^{M-1}\left[\Psi_{1D}(I_k) \prod_{j>k} (p_k-p_j)^{n_j}\right] 
\label{fig:box2d}
\end{equation}
where the prefactor is a nonnegative function.
Similar expression can be written down by factorizing along the lines $q=q_i$.
The wave function therefore
factorizes in a manner very similar to  the harmonic oscillator
and periodic fermion gas. Therefore the proof of the two nodal domains
 above can be constructed in a similar manner.
First, it is not too difficult to show that
the there are only two domains in the system with three particles.
Next, one assumes that this is the 
case for a system with $M$ lines and then it is straightforward to show
 that the same applies to the system with $M+1$ 
lines.
Note that Eq.\ref{fig:box2d} suggest
that apparently the wave function 
has  both rotation and translation invariance. This is strictly not true,
due to the nonlinearity of the map $(p,q) \to (x,y)$.
What is true, however, is that node is not crossed during 
rotations and translations since  $(x,y)\to (p,q)$ is homeomorphic.
Therefore rotations and translations 
change the wave function values but not the sign since
there is no reordering of either the lines or the particles along the lines.
That enables us
to use rotations and translations to prove the 
connectedness assuming that we take additional care
 to keep the paths contained 
within the box. 
The proof then follows similar line of arguments as presented before.
It is easy to show that the three-particle state has only two nodal cells.
Place the particles on a lattice and position the system into the center of the box.
We assume that
the lattice constant is sufficiently 
small
so that translations by one lattice constant 
would not push the particles out of the box. 
It is then straightforward to show that the four translations 
$T^x_{-a}T^y_{-a}T^x_aT^y_a$
exchange three particles at the edge of the system and 
by similar exchanges one can show connectedness
of all particles for arbitrary size. Therefore the ground state closed-shell wave functions
for particles in a box have the same
generic properties as the fermionic models studied in preceding sections.

\section{\label{sec:theorem}
 IV. Minimal number of nodal cells theorem for spin-polarized noninteracting systems.} 
Using the results and proofs derived in this section
and  also in the previous paper \cite{mitasshort} we can write
the following theorem.

{\it Theorem.} Consider noninteracting or mean-field spin-polarized 
fermions in $d>1$  with an exact wave function given by a 
Slater determinant of one-particle states times a prefactor which does affect
the fermion nodes.
 Let the one-particle states are  
such that the Slater matrix
elements can be rearranged to monomials in particle coordinates or in coordinates transformed
by a homeomorphic map. Then for an arbitrary size closed-shell ground state 
the corresponding 
wave function has the minimal
number of two nodal domains.

The theorem covers a number of paradigmatic models and it is quite suggestive to think about this 
as being 
 related to general mathematical properties of zeros of functions defined through
determinants. In fact, the 
factorizations which enabled us to prove the two nodal cells property is directly
related to the properties of multiple hyperplane configurations and to the multi-variate  
Vandermonde  determinant theorem which can be found in mathematical literature in
various contexts  \cite{varchenko, lagr}.

Considering that we restricted the proofs to noninteracting systems and we relied
on the fact that the matrix elements are monomials, it is useful to consider
cases
which go beyond such a framework. 
This line of thought leads us to the following interesting questions:

i) Is the {\it two nodal cell property} valid for noninteracting cases with Slater 
{\it matrix elements
   not reducible to monomials} ?

ii) What is the {\it impact of interactions} ?

A tentative answer to the first question is given in the next section where we show that
atomic states in Coulomb potential, which cannot be reduced to monomials due to the
shell structure, exhibit the same property.  We will not investigate the impact of interactions
for spin-polarized systems in this  
paper. However, we will study and prove the two nodal cells for 
 perhaps even more important cases
of interacting spin-unpolarized systems in the 
section VI.

\section{\label{sec:spdsp} V. Spin-polarized atomic states.} 
In the previous paper \cite{mitasshort}
 we have proved that the atomic
spin-polarized state $1s2s2p^3$ has two nodal cells for both 
noninteracting and HF wave functions. The proofs for 
atomic states are more 
involved  since for Coulomb potential it is more
difficult to find appropriate factorizations.
The main complication is that orbitals in 
different subshells and 
angular momentum channels have, in general, 
different radial dependences which cannot
be all factorized out of the determinant into a common prefactor. 
Nevertheless, it is possible to demonstrate the two nodal cells for atomic 
states for several spin-polarized (half-filled) main subshells 
(and possibly, for all the states relevant
for the periodic table of elements). We will
 illustrate the idea of the proof on
the spin-polarized $^{15}S(1s2s2p^33s3p^33d^5)$ state with 14 electrons 
and then point out 
how the proof can be extended
to larger systems.
The one-particle orbitals
are $\rho_{1s}(r),$ $\rho_{2s}(r),$ $\rho_{2p}(r)x,$
$\rho_{2p}(r)y,$ $\rho_{2p}(r)z$,...,$\rho_{3d}(r)(2z^2-x^2-y^2)$, etc,
and we use dimensionless coordinates which are rescaled as
${\bf r}\leftarrow Z{\bf r}/a_0$ with $Z$ being the nuclear
charge and $a_0$ the Bohr radius.
The wave function is given by
\begin{equation}
\Psi_{at}(1,...,14)=\nonumber
\end{equation}
\begin{equation}
{\rm det}\{\rho_{1s}^{*},\rho_{2s}^{*},x,y,z,
\rho_{3s}^{*}, \phi_{3px}^{*},..., \phi_{3dz^2}^{*}, ...\}
\end{equation}
where $\rho_{1s}^{*}(r)=\rho_{1s}(r)/\rho_{2p}(r)$,
 $\rho_{2s}^{*}(r)=\rho_{2s}(r)/\rho_{2p}(r)$ and
$\phi_{3px}^{*}({\bf r})=x\rho_{3p}(r)/\rho_{2p}(r)$, ..., $\phi_{3dz^2}^{*}({\bf r})
=\rho_{3d}(r)(2z^2-x^2-y^2)/\rho_{2p}(r)$, ...,
etc,  where
we factorized the
 nonnegative radial function $\rho_{2p}(r)$
out of the determinant.

We will show the connectedness of all the particles in two steps. 
We first demonstrate that the Slater determinant can be factorized 
into subdeterminants corresponding to subshells $1s$,  
$2s2p^3$ and $3s3p^33d^5$. That will enable us to show
that the particles within each subshell
are connected. In the second step we show, that the particles can exchange
between the subshells.
For this purpose we will specify the explicit paths
and carry out a straightforward numerical check that there is no node crossing
by  tracing the wave function along
the paths. 

In order to show the factorization into shells
we position the particles as follows: particle 1 
is in the origin, particles 2 to 5 are on the surface
of a sphere with the
radius $\eta_a$ and particles 6 to 14 are 
are on the surface
of a sphere with the radius $\eta_b$. 
The radii $\eta_a$ and $\eta_b$ are given by the radial nodes of 
the orbitals $\rho_{2s}(r)$ and $\rho_{3p}(r)$, 
ie, $\rho_{2s}(\eta_a)=0$ and  $\rho_{3p}(\eta_b)=0$. 
The Slater determinant can be written in the form
\begin{equation}
\Psi_{at}(1,...,14)=
{\rm det}
\left[ \begin{array}{ccc}
{\bf A } & {\bf D } &{\bf  G} \\
{\bf B } & {\bf E' } & {\bf E}\\
{\bf C } & {\bf F } & {\bf  H}\\
\end{array} \right] 
\label{eq:spd}
\end{equation}
where the block matrices ${\bf A}$ to ${\bf H}$ are given as follows
\begin{equation}
{\bf A}=\left[ \begin{array}{ccccc}
\rho_{1s}^{*}(0)   &  a_1       &  a_1   & a_1  & a_1 \\
\rho_{2s}^{*}(0)   &  0       &  0   & 0  & 0 \\
0 &  x_2      & x_3 & x_4  &  x_5 \\
0 &  y_2     & y_3  & y_4  &  y_5 \\
0 &  z_2     & z_3 & z_4 &  z_5 \\
\end{array} \right]  
\end{equation} 
\begin{equation}
{\bf B}=\left[ \begin{array}{ccccc}
\rho_{3s}^{*}(0)   &  a_3       &  a_3   & a_3  & a_3   \\
0 &  \phi_{3px}^*(2)      &  ...  & 
 &  \phi_{3px}^*(5) \\
0 &  \phi_{3py}^*(2)      & ...  &  &  \phi_{3py}^*(5)\\
0 &  \phi_{3pz}^*(2)       &  &  &  \phi_{3pz}^*(5) \\
\end{array} \right]    
\end{equation}   
\begin{equation}
{\bf C}=\left[ \begin{array}{ccccc}
0 &  \phi_{3dz^2}^*(2)      &  ...  & &  \phi_{3dz^2}^*(5) \\
0 &  \phi_{3dx^2}^*(2)      & ...  &  &  \phi_{3dx^2}^*(5)\\     
0 &  \phi_{3dxy}^*(2)       & ...  &  &  \phi_{3dxy}^*(5) \\
0 &  \phi_{3dyz}^*(2)       & ... &  &  \phi_{3dyz}^*(5) \\
0 &  \phi_{3dxz}^*(2)       & ... &  &  \phi_{3dxz}^*(5) \\
\end{array} \right]       
\end{equation}   
\begin{equation}      
{\bf D}=\left[ \begin{array}{cccc}
 b_1       &  b_1   & b_1  & b_1   \\
 b_2       &  b_2   & b_2  & b_2   \\
x_6 &  ...        & ...  &  x_9 \\
y_6 &  ...        & ...  &  y_9 \\
z_6 &  ...        & ...  &  z_9 \\
\end{array} \right]
\end{equation}
\begin{equation}
{\bf E}=\left[ \begin{array}{ccccc}
 b_3       &  b_3   & b_3  & b_3 & b_3  \\
0 &  ...        &  & ... &  0 \\ 
0 &  ...        &  & ... &  0 \\   
0 &  ...        &  & ... &  0 \\
\end{array} \right] 
\end{equation} 
\begin{equation}
{\bf F}=\left[ \begin{array}{cccc}
  \phi_{3dz^2}^*(6)      &  ...  & &  \phi_{3dz^2}^*(9) \\
  \phi_{3dx^2}^*(6)      & ...  &  &  \phi_{3dx^2}^*(9)\\
  \phi_{3dxy}^*(6)       & ...  &  &  \phi_{3dxy}^*(9) \\
  \phi_{3dyz}^*(6)       & ... &  &  \phi_{3dyz}^*(9) \\
  \phi_{3dxz}^*(6)       & ... &  &  \phi_{3dxz}^*(9) \\
\end{array} \right]
\end{equation}
\begin{equation}          
{\bf G}=\left[ \begin{array}{ccccc}
 b_1       &  b_1   & b_1  & b_1 & b_1  \\
 b_2       &  b_2   & b_2  & b_2 & b_2  \\
x_{10} &  ...        & ...  &  ... & x_{14}\\
y_{10} &  ...        & ...  &  ... & y_{14}\\
z_{10} &  ...        & ...  &  ... & z_{14}\\
\end{array} \right]
\end{equation}
\begin{equation}
{\bf H}=\left[ \begin{array}{ccccc}
  \phi_{3dz^2}^*(10)      &  ...  & &  &  \phi_{3dz^2}^*(14) \\
  \phi_{3dx^2}^*(10)      & ...  &  & &  \phi_{3dx^2}^*(14)\\
  \phi_{3dxy}^*(10)       & ...  & &  &  \phi_{3dxy}^*(14) \\
  \phi_{3dyz}^*(10)       & ... &  & &  \phi_{3dyz}^*(14) \\
  \phi_{3dxz}^*(10)       & ... &  & &  \phi_{3dxz}^*(14) \\
\end{array} \right]
\end{equation}
and $a_1=\rho_{1s}^{*}(\eta_a)$, $a_3=\rho_{3s}^{*}(\eta_a)$,
$b_1=\rho_{1s}^{*}(\eta_b)$, $b_2=\rho_{2s}^{*}(\eta_b)$, $b_3=\rho_{3s}^{*}(\eta_b)$.
The block matrix ${\bf E'}$ is the same as  ${\bf E}$ except that it has only four columns 
instead of five. 

The following three row additions 
in the Slater matrix given by Eq.\ref{eq:spd} 
allow for factorization into subdeterminants.
We first eliminate elements $a_1$ in the first row   
by adding an appropriate multiple of the fifth row. Similarly, 
we eliminate the terms with $b_1$ in the first row  
by adding a multiple of the second row. This
leads to the first row having only one nonzero element
$A'_{11}=\rho_{1s}^{*}(0)$
$-a_1\rho_{3s}^{*}(0)/a_3$
$-(b_1-a_1b_3/a_3)\rho_{2s}^{*}(0)/b_2$ and reduces the Slater matrix
by the first row and the first column.
Finally,
 by adding a multiple of the second row to the fifth row, 
both ${\bf E'}$ and ${\bf E}$ become zero matrices and we can write
\begin{equation}
\Psi_{at}(1,...,14)=A'_{11}\, {\rm det} {\bf B'}\,
{\rm det}
\left[ \begin{array}{cc}
 {\bf D' } &{\bf  G'} \\
{\bf F } & {\bf  H}\\
\end{array} \right] 
\end{equation}
where  ${\bf B'}$ is the matrix ${\bf B}$ without the first column,
while ${\bf D' }, {\bf  G'}$ are the matrices ${\bf D}, {\bf  G}$ without the first row.
The three factors in the expression above correspond to the three main 
subshells namely, $1s$, $2s2p^3$ and $3s3p^33d^5$ with the dependence
on positions of particles 1, 2-5 and 6-14,
respectively.
The key point is that each of the subshells represents a system of particles on a sphere in
the $S$-symmetry state and for such cases 
we proved that the particles are connected.
That concludes the argument that the particles within each subshell are connected.

\begin{figure}[ht]
\centering
\includegraphics[width=2.0in,clip]{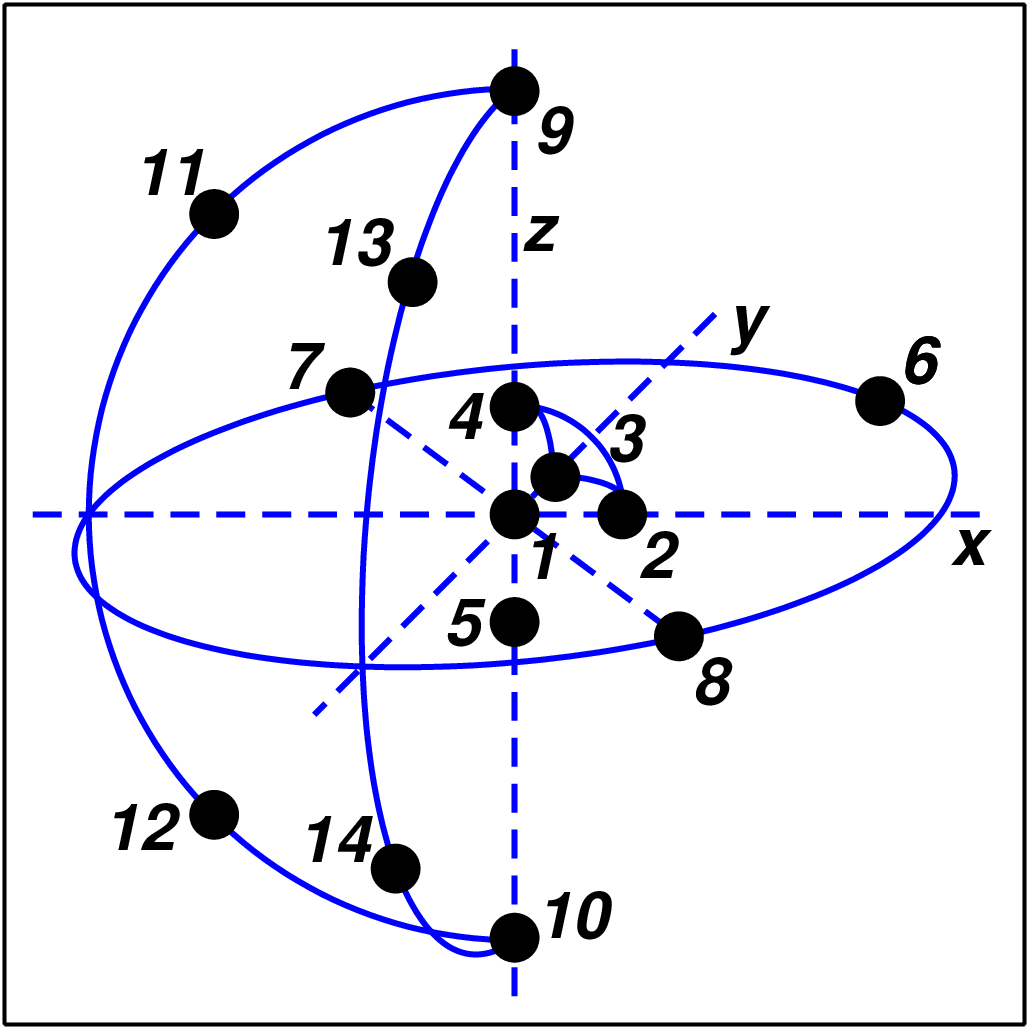} 
\caption{Positions of 14 electrons for the proof that triple exchanges connect the particles 
in the $^{15}S(1s2s2p^33s3p^33d^5)$  state.
Particles 2 to 5 lie on the spherical surface with 
the radius equal to the radial node of $\rho_{2s}(r)$ orbital.
Particles 6 to 14 lie on the spherical surface with the radius equal to the radial node of $\rho_{2p}(r)$.}
\label{fig:3spd}
\end{figure}

\begin{figure}[ht]
\centering
\includegraphics[width=2.4in,clip]{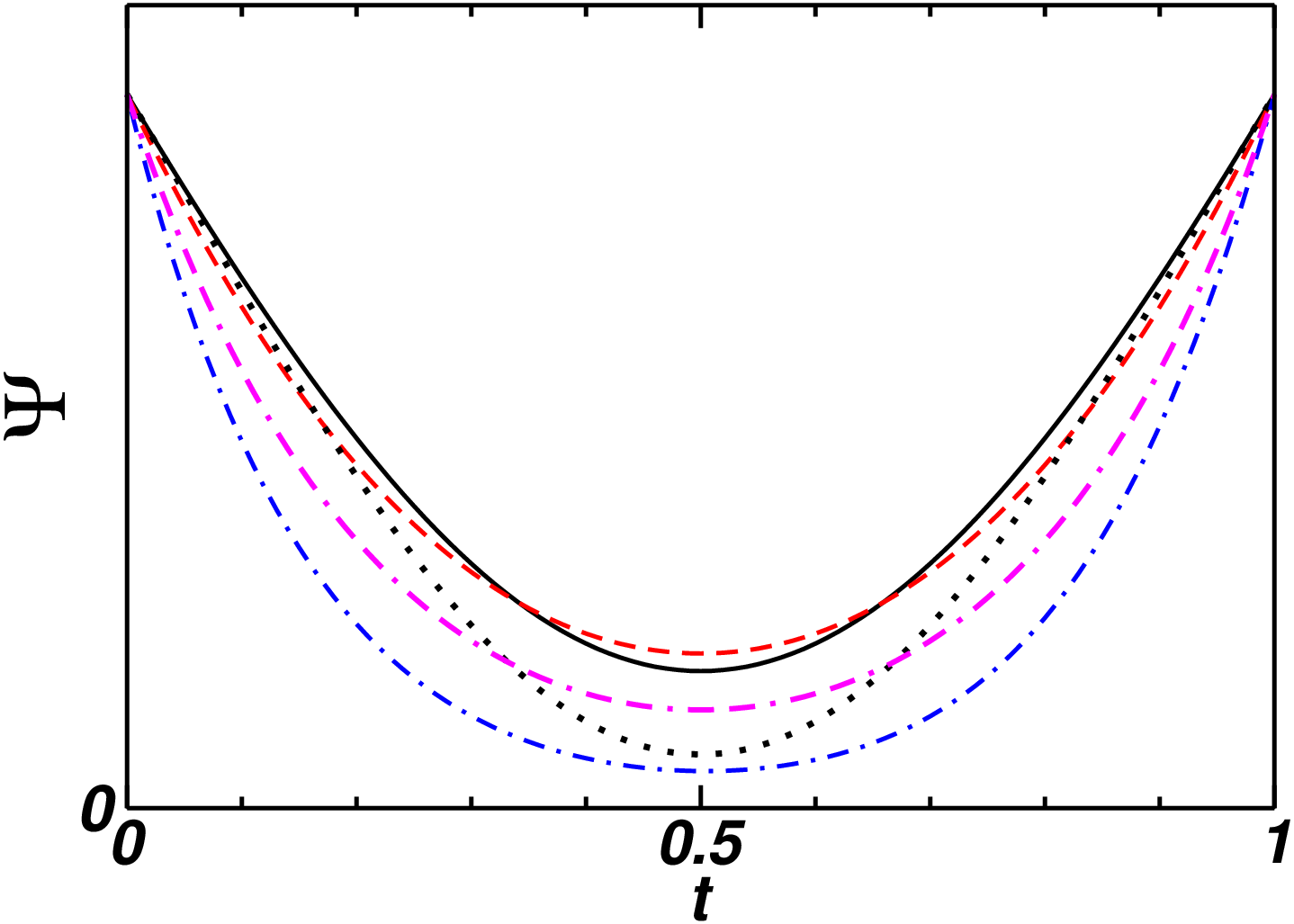}
\caption{
Wave function of the $^{15}S(1s2s2p^33s3p^33d^5)$ 
state (arb. u.) along triple exchange paths parametrized by $t$. Particles are positioned 
as given in Fig.~\ref{fig:3spd}. The plotted 
exchanges:  1,2,3 (full line); 2,3,4 (dashed line); 2,3,6 (dotted line); 6,7,9 (dashed-dotted line); 
9,11,13 (double dashed-dotted line). By symmetry, these non-crossing exchanges connect all the particles of this 
state.}  
\label{fig:atexch}
\end{figure}

More difficult part of the proof is to show that one can exchange the particles between the subshells.
We were able to prove analytically the exchange between $1s$ and $2s2p^3$ in our previous paper. For
larger cases the analytic proof becomes very tedious and it is much more efficient
to evaluate numerically the determinant along the following 
exchange paths.
The particles are positioned as sketched in Fig.\ref{fig:3spd}
with the coordinates given by
${\bf r}_1=(0,0,0)$, ${\bf r}_2=(\eta_a,0,0)$, ${\bf r}_2=(0,\eta_a,0)$,
${\bf r}_{4,5}=(0,0,\pm \eta_a)$, ${\bf r}_6=(\eta_b/\sqrt{2},\eta_b/\sqrt{2},0)$,
${\bf r}_{7,8}=(\pm\eta_b/\sqrt{2},\mp\eta_b/\sqrt{2},0)$, ${\bf r}_{9,10}=(0,0,\pm\eta_b)$,
${\bf r}_{11,12}=(\eta_b/\sqrt{2},0,\pm\eta_b/\sqrt{2})$ and ${\bf r}_{13,14}=(0,\eta_b/\sqrt{2},\pm\eta_b/\sqrt{2})$.
There exists enough of triple 
exchanges of neighbouring particles, along the sides of corresponding triangles, to connect 
all the particles into a single cluster. 
For example,
the wave function values for exchanges $123\to 231$, $236\to 362$ (and other exchanges for 
illustration) are given in Fig.~\ref{fig:atexch}.

This is enough to show that the three subshells are connected
since many other exchanges are identical due to the symmetries in particle positions 
and the wave function $S$ symmetry. 
For the illustration in Fig.~\ref{fig:atexch} we used the noninteracting radial orbitals.
 For HF orbitals one gets the same qualitative picture since
the basic spatial properties of the noninteracting
and HF orbitals are qualitatively the same 
(ie, ordering of radial nodes of $\rho_{nl}(r)$ orbitals,
behaviour at nucleus, tails,
 etc). The proof uses only the fact that some of the radial nodes of the one-particle orbitals
are ordered as in the non-interacting case so it can be extended to HF wave functions as well. 

One can expand the proof to larger systems, such as for the state with
occupied fourth main subshell  
 $4s4p^34d^54f^7$ and beyond. The factorization is similar: the particles in the
fourth subshell are positioned on the spherical surface with the radius
equal to the radial node of $\rho_{4d}(r)$ orbital. Somewhat long
but straightforward
algebraic rearrangements show
that the Slater determinant of the $30\times 30$ matrix can be 
reduced to product of subdeterminants corresponding to the subshells.
For the purpose of this paper we do not deem necessary to go through the explicit proof.

\section{\label{sec:hegsu} VI. Interacting spin-unpolarized fermions.}

It is straightforward to understand 
the fermion nodes of noninteracting
spin-unpolarized or partially polarized systems with more than one-electron in each
spin-channel.
The wave function is a product of spin-up and -down Slater determinants and the 
number of nodal cells is the product of the number of cells in each subspace.
(As mentioned before, we assume that the Hamiltonian, even with interactions considered
below, does not include terms with spin so that particles can be assigned 
to the spin subspaces.) For the ground
states with two nodal cells in each subspace we get $2\times2=4$ nodal cells.
This is the nodal structure of a single configuration Hartree-Fock wave function
and  as such has been used in many 
fixed-node QMC calculations. However, analyses of small interacting systems
revealed that this is
not correct and interactions do change the nodal topologies
and the number of nodal cells. 
For the first time 
this has been demonstrated 
for the Be atom \cite{dariobe}
and then also for a few other atoms and small molecules \cite{cmt28,bajdichpfaffians}.

In the previous paper \cite{mitasshort} we have outlined a proof 
showing the two nodal domains property for $2D$
harmonic fermions in a closed-shell singlet state for arbitrary size. The most interesting feature
of the proof was that the result was very 
robust in the sense that almost any arbitrary weak interaction would induce the change in the 
topology of the nodal surfaces. 

We
refresh some of the key notions and arguments here. 
Let us assume a spin-unpolarized system in its
closed-shell singlet ground state of $2N$ particles.
Consider a simultaneous exchange of an odd number of spin-up pair(s)
{\it and} an odd number of spin-down pair(s).
For noninteracting wave functions such simultaneous pair exchanges
imply that the node will be crossed once or multiple times. This must be the case
whenever the spin-up and -down subspaces are independent of each other, such as 
in the mean-field and HF wave functions. However,
if there exists a point $R_f$ such that during the simultaneous
spin-up and -down pair exchanges the inequality
   $|\Psi|>0$ holds along the whole path,
 then the wave function has only two nodal cells.

Using this property we can now demonstrate the two nodal cell for several types of systems.
Under rather general conditions, 
 we will show that the correlation included in
the Bardeen-Cooper-Schrieffer (BCS) pairing
wave function \cite{sorella,carlson} given by
\begin{equation}
\Psi_{BCS}(1,...,2N)={\rm det} [\Phi(i,j)] 
\label{eq:bcs}
\end{equation}
smoothes out 
 the noninteracting four nodal cells
into the minimal two. 
Here $\Phi(i,j)=\Phi(j,i)$ is a singlet pair orbital for
$i\negthinspace\negthinspace\uparrow $ and
$j\negthinspace\negthinspace\downarrow$ fermions and we decompose it
into noninteracting and correlated components
$\Phi(i,j)= \Phi_{0}(i,j) + \Phi_{corr}(i,j)$. Using one-particle orbitals
we can write 
$\Phi_{0}(i,j)=\sum_{\alpha}\negthinspace \phi_{\alpha}(i)\phi_{\alpha}(j)$ where
the sum is over HF (or noninteracting) orbitals 
while $ \Phi_{corr}(i,j)=\sum_{\alpha\beta}c_{\alpha\beta}\phi_{\alpha}(i)\phi_{\beta}(j)$
where $\{c_{\alpha\beta}=c_{\beta\alpha}\}$ are variational parameters and 
the sum is over unoccupied or virtual ("correlating") 
orbitals. The BCS wave function was originally introduced for conventional
superconductors, however, it proved to be very successful also in describing 
the correlation effects in electronic structure problems \cite{sorella,bajdichpfaffians}.   

\subsection{VI.a. $3D$ harmonic spin-unpolarized fermions with interactions.}

\begin{figure}[ht]
\centering
\includegraphics[width=1.8in,clip]{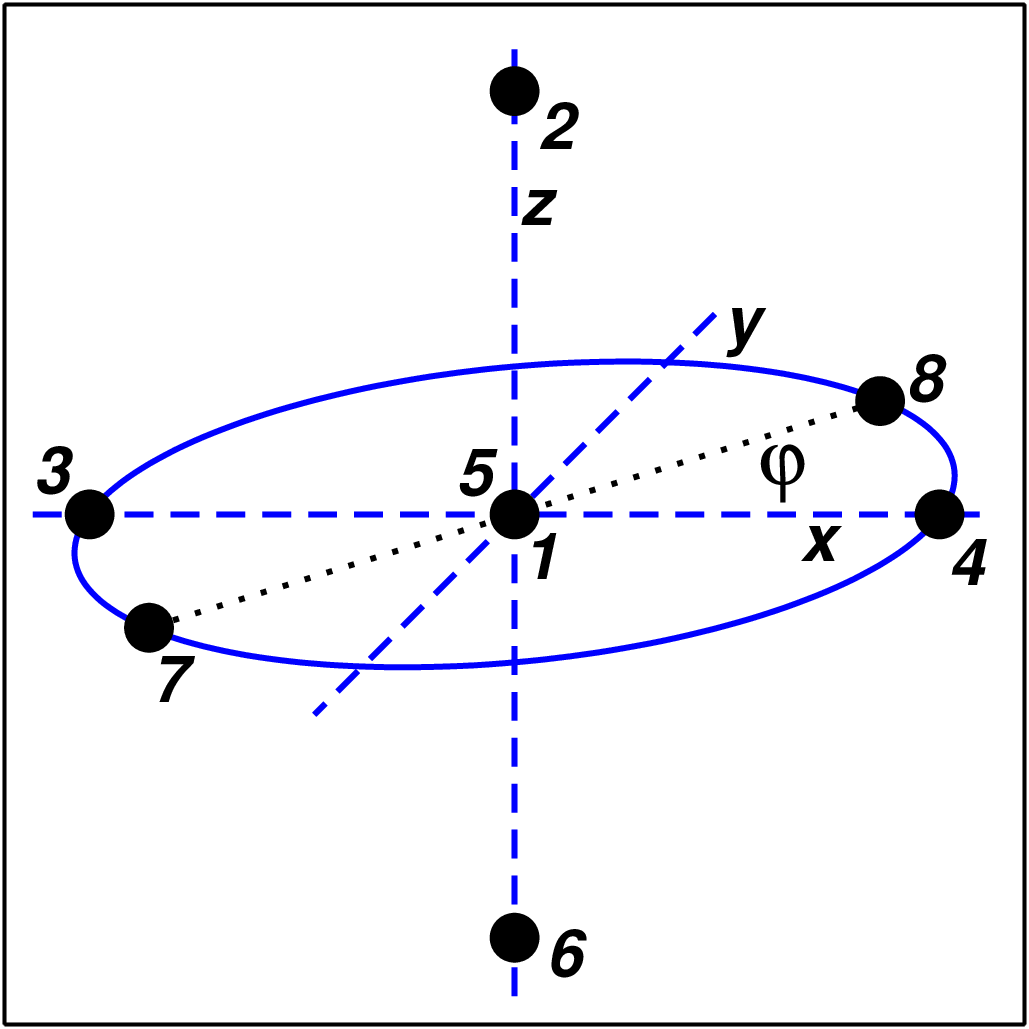}
\caption{Positions of $2N=8$ fermions of the $3D$ harmonic well system. Fermions 1 to 4 have spins up,
while 5 to 8 have spins down. Fermions 1 and 5 are at the origin while the rest of the particles
lie on the surface of the sphere with $r=1$. Note that rotation of the system
by $\pi$ around the $z-$axis exchanges particles $3,4$ and $7,8$.
See the text for details. }
\label{fig:3dho}
\end{figure}

For the $3D$ harmonic oscillator one can show that the interactions lead to the
minimal number of nodal cells in a rather simple and elegant way.
  First let us consider
a small system which is easy to analyze. We illustrate
the idea on $2N=8$ particles in the singlet
ground state with the particle positions given in Fig.~\ref{fig:3dho}. 
The corresponding closed-shell
singlet ground state for 
the $3D$ harmonic oscillator is $^1 S(1s^22p^6)$. (Note that for the harmonic potential
the $2p$ state is below the $2s$ state,
unlike for the Coulomb potential.)
For simplicity, we drop the gaussian prefactors since they do not affect
the nodes and 
the pairing 
functions can be then written
as follows
\begin{equation}
 \Phi_{0}(i,j) = 1 + 3{\bf r}_i\cdot {\bf r}_j
\end{equation}
and 
\begin{equation}
 \Phi_{corr}(i,j) = \gamma [3({\bf r}_i\cdot {\bf r}_j)^2 -r_i^2r_j^2]
\end{equation}
where the noninteracting part is constructed from the occupied $1s,2p$
 orbitals while for the correlating 
part the unoccupied $3d$ subshell orbitals were used. 
The constant $\gamma$ is a variational parameter.
Clearly, both the pair orbitals
and the wave function are spherically symmetric.
Assuming the positions as given in the Fig.~\ref{fig:3dho},  which are of high symmetry
and therefore simplify the evaluation of the wave function, we get 
\begin{equation}
\Psi(1,...,8)= 4\gamma\cos\varphi[1-(3+\gamma)\cos^2\varphi]/3
\end{equation}
where $r_1=r_5=0$,  $\; r_i=1, i=2,3,4,6,7,8$
and $\cos\varphi={\bf r}_4\cdot {\bf r_8}$.

For $\gamma =0$ the wave function vanishes since the particles in both
spin subspaces lie on the noninteracting node (ie, four particles on the same plane). 
The key point is that for
$\gamma\neq 0$ and $\varphi\neq \pi/2$ the wave function
does not vanish.
Since the wave function is spherically symmetric we can rotate the particles 
around the $z-$axis by $\pi$. This transformation 
exchanges particles 3 and 4 in the spin-up 
channel and particles 7 and 8 in the spin-down channel. 
 The wave function is rotationally
invariant and nonvanishing what clearly implies that the BCS wave function has
spin-up and -down subspaces interconnected since simultaneous exchanges in the spin-up  
and spin-down {\it does not } hit the node. 
On the other hand, it is easy to check
that the rotation of the particles in one spin channel only causes a node-crossing
since then $\cos\varphi$ vanishes either at $\pi/2$ or $3\pi/2$. 

It is clear that our argument is correct
whenever the strength of the interaction
is small so that the BCS wave function is accurate enough. 
The variational 
parameter $\gamma$ is related to the interaction strength. This
 implies that the topological 
change from two to four nodal cells takes place for arbitrary small, but 
nonvanishing, interaction strength.
The only assumptions were
that the interaction will induce electron correlation and lead to a nonzero
variational parameter $\gamma$ (or, in general, to a nonvanishing correlation
component) and that the wave function is spherically
symmetric. Our analytic argument therefore shows
that for weak
interactions the nodal cell "degeneracy" is lifted
and the multiple cells are smoothed to the minimal
number of two. 

It is important to note that
one can find different pictures in other circumstances. For example,
 for strong or nonlocal interactions, imposed additional symmetries or
large degeneracies, the nodal changes can be different and the resulting
nodal topology might exhibit more than the two nodal cells;
this aspect is further
discussed in the conclusion section and require more investigation. 

Coming back to our example of
harmonic fermions with interactions, using similar arguments 
the two nodal cells can be demonstrated for
an arbitrary size closed-shell ground state. 
Consider $2N$ particles in a singlet closed-shell ground state.
The closed-shell states can be labeled by $M$ which represents the "Fermi momentum"
for the harmonic oscillator and $N=(M+1)(M+2)(M+3)/6$.  Let us
decompose $N$ into the maximum 
odd number of 
pairs $N_P$ and the rest. Therefore we write  
$N=2N_P+J$ where $N_P$ is an odd integer while $J$ has one of the values $J=0,1,2,3$, so 
that we can also define $J=N\, {\it mod}\, 4$.
For example, in
the previous example with $2N=8$ particles we have $N_P=1$ and $J=2$. First, we place $J$
particles from each spin channel on the $z$-axis in
  distinct positions.  Next,
we form $N_P$ pairs 
 in each spin subspace
so that, say, the pair $i\negthinspace\uparrow,
(i+1)\negthinspace\uparrow $ is positioned as  
given by $(r_i,\vartheta_i,\varphi_i)=(r_k,\vartheta_k,\varphi_k)$,
 $\,(r_{i+1},\vartheta_{i+1},\varphi_{i+1})=(r_k,\vartheta_k,\varphi_k+\pi)$,
where $r,\vartheta,\varphi$ are the spherical coordinates.
Here $k=1,...,N_P$ labels the pairs
in the spin-up channel; the spin-down particles are 
placed similarly and labeled by the pair index $l=N_p+1,...,2N_P$.
In this configuration the particles lie on the 
noninteracting node since 
\begin{equation}
 {\rm det}\left[\Phi_0(i,j)\right]={\rm det}\left[\sum_{n+m\leq M} \phi_{nm}(i)
 \phi_{nm}(j)\right]=
\nonumber
\end{equation}
\begin{equation}
={\rm det}\left[\phi_{nm}(i)\right]
{\rm det}\left[\phi_{nm}(j)\right]=\Psi_{HF}^{\uparrow}\Psi_{HF}^{\downarrow}
\end{equation}
The rotation of the system by $\pi$ causes node-crossings 
in both spin channels
so that both spin-up and -down Slater determinants must  vanish due to the rotational
 invariance.
Now, if all the $N_P$ pair distances and angles
$r_k,\varphi_k,\vartheta_k,r_l,\vartheta_l,\varphi_l$ are distinct, 
then each of the matrices $\left[\phi_{nm}(i)\right], \left[\phi_{nm}(j)\right]$
has exactly one linearly dependent row, ie, both have ranks $N_P-1$.
This can be verified directly
for small values of $M$ and then using induction for any $M$.
Consequently, the matrix $\left[\Phi_0(i,j)\right]$ has linear dependence
in one row and one column, ie, it has the rank of $N_P-1$ as well. In general,
adding virtual states through $\Phi_{corr}(i,j)$ provides
independent rows/columns
which eliminate the linear dependency so
that ${\rm det}\left[\Phi_0(i,j)+\Phi_{corr}(i,j)\right]$ is nonzero. Let us now assume 
that the interactions do not break the rotation invariance. Since at this point the
wave function does not vanish and is rotationally invariant,  
it does not vanish for the whole exchange path implying that the correlated 
BCS wave function has only two nodal cells regardless of the size $2N$.

 Clearly, the assumption of the rotational invariance for the interacting
case might be too
restrictive since the interactions could possibly break the spherical symmetry.
To demonstrate  the two nodal cells in such a case one would need to show
 that the wave function does not vanish for the complete exchange path.

\subsection{VI.b. $2D$ harmonic spin-unpolarized fermions with interactions for 
$N=2N_P+J$ when $J=2,3$.}

\begin{figure}[ht]
\centering
\begin{tabular}{ccc}
\includegraphics[width=1.4in,clip]{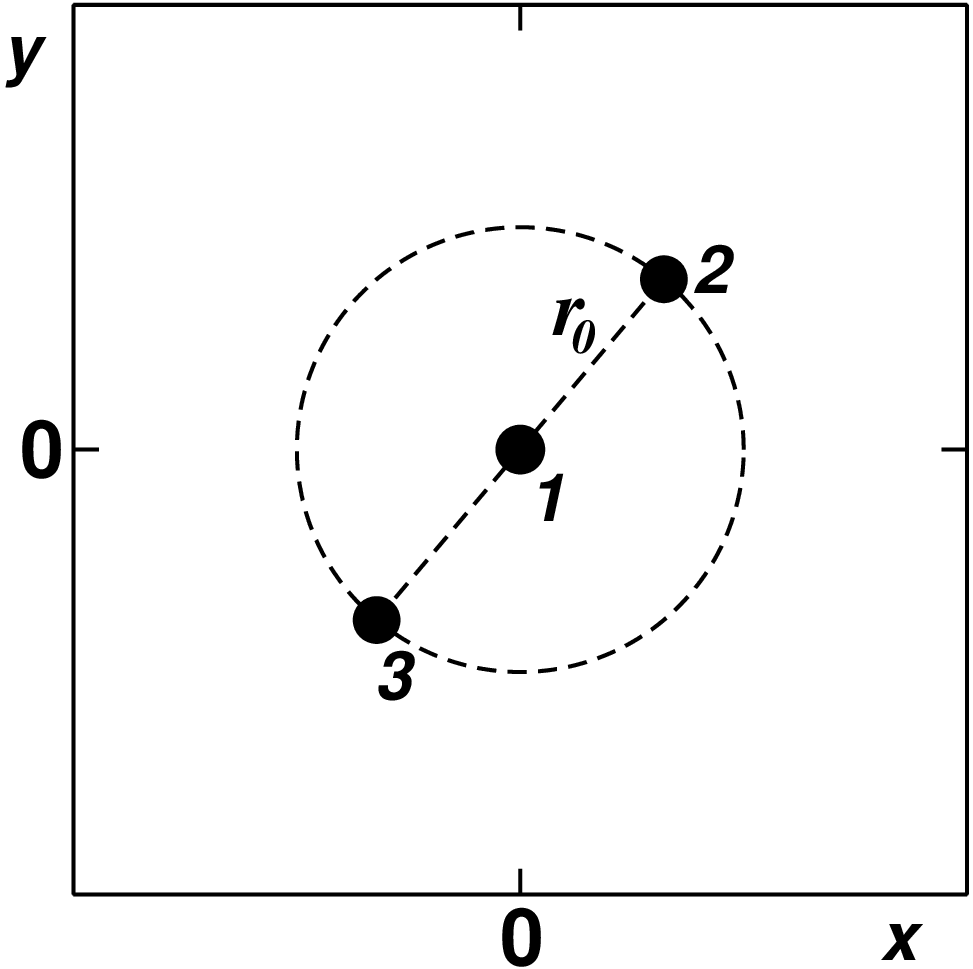} & \quad &
\includegraphics[width=1.4in,clip]{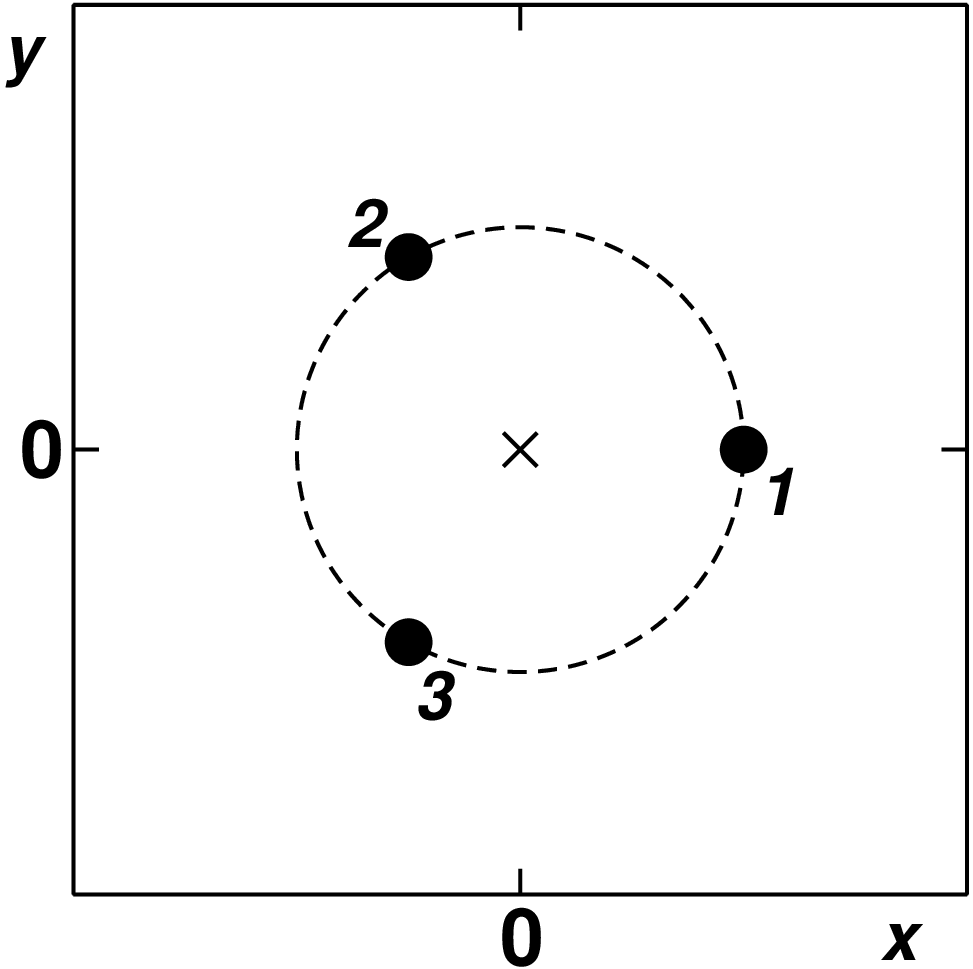} \\
(a) & \quad & (b) \\
 \quad & \quad & \quad \\
 \quad & \quad & \quad \\
\includegraphics[width=1.4in,clip]{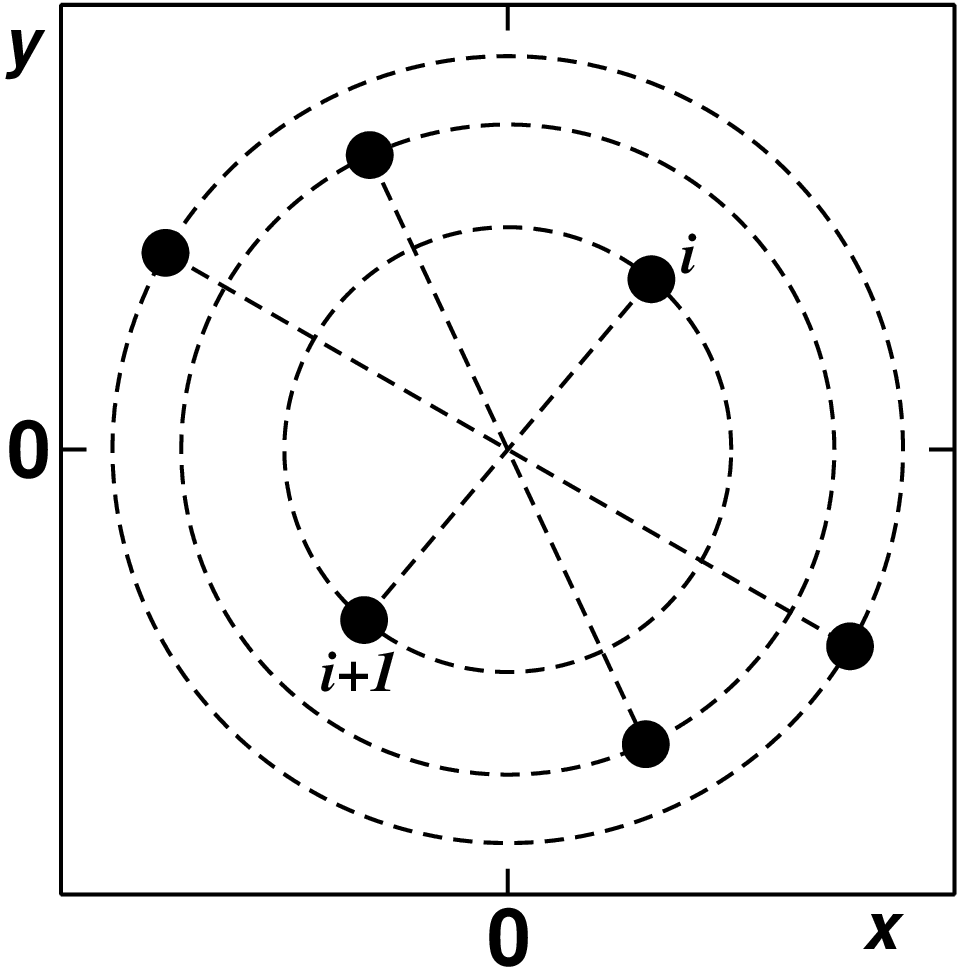} & \quad &
\includegraphics[width=1.4in,clip]{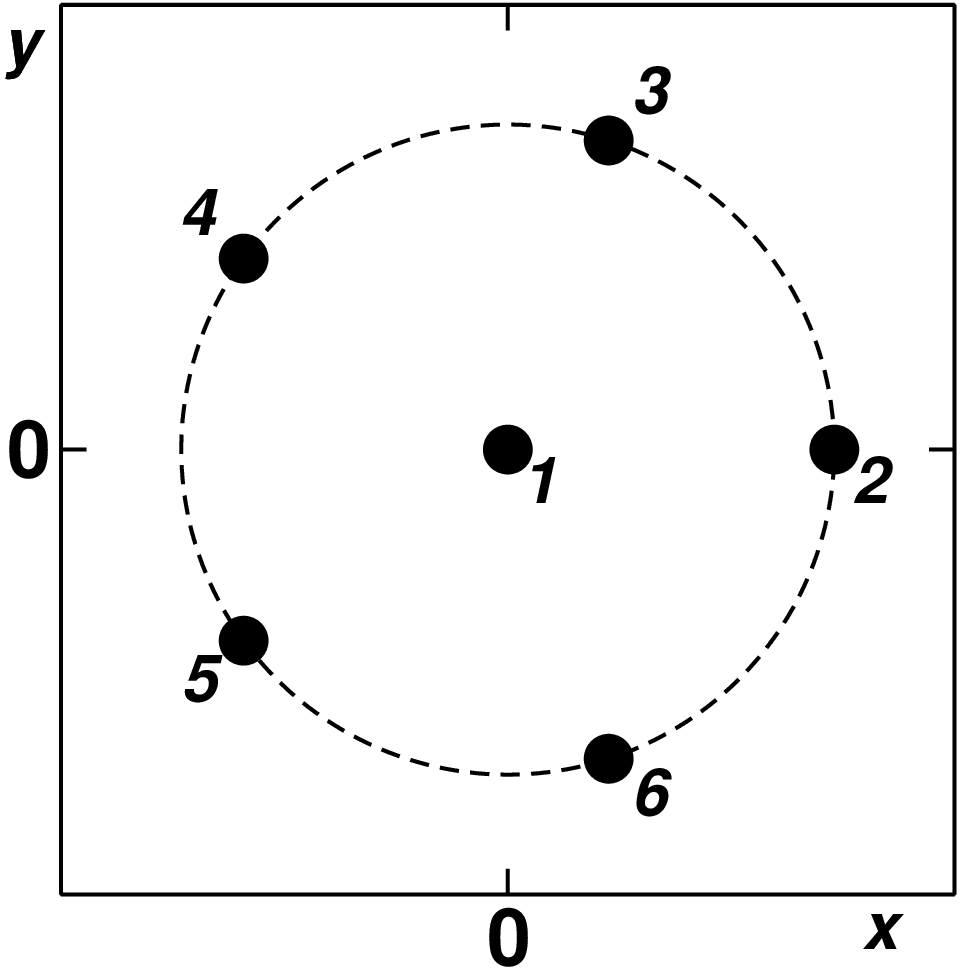} \\
(c) & \quad & (d) \\
\end{tabular}
\caption{Alternative positions of three and six spin-polarized fermions for the $2D$ harmonic
potential.
(a) Three particles positioned on the noninteracting node.
(b) Three particles in positions with a nonvanishing wave function.
(c) Six particles positioned on the noninteracting node.
(d) Six particles in positions with a nonvanishing wave function.
See the text for details. }
\label{fig:2dho}
\end{figure}

In our previous paper we have demonstrated
the two nodal cells for the
ground states with 
the number of particles $2N$ where $N=2N_P+J$ and $J=0,1$. (We use the 
same notation as in $3D$ where $N_P$ is the maximum odd number of pairs and $J=0,1,2$ or 3).
Note that the states with $J=0$ or 1 exist at any size. Here we would like to present 
an alternative
and a little bit longer
proof which covers $J=0,1$ but also remaining cases
 when $J=2,3$, eg, 
for $N=21=2\times9 +3,\;$ $N=28=2\times13 +2$, etc.
In $3D$, the rotation axis can accommodate multiple particles so that one can always
form an odd number of pairs for the exchanges. In $2D$, the
symmetry point (origin) can accommodate only one particle   
from each spin subspace therefore the proof has to be modified.
For this purpose we outline a factorization
which is different from the previous paper \cite{mitasshort}. Let us remind that for $2D$
harmonic oscillator the closed-shell states and the system size are labeled by
$M=1,2, ...$ where $n+m\le M$ with the number of fermions in one spin
channel given by $N=(M+1)(M+2)/2$.  We express the one-particle states 
as polynomials in $r^2$ and $(re^{i\varphi})^m$  where $r, \varphi$ are 
the cylindric coordinates (we again omit the gaussian factors
which are irrelevant for the nodes).  Note that for a given $M$ the 
quantum number $m$ increases or decreases by 2 unlike
in $3D$. Therefore for $M=2$ we have states $re^{\pm i\varphi}$, for $M=3$ the 
additional states are $(r^2-1), r^2e^{i\pm 2\varphi}$, etc. 
We first write down the wave functions for three- and six-particle 
spin-polarized systems with positions given in Fig. \ref{fig:2dho}.

The three particle wave function is given by ${\rm det}\{1,re^{\pm i\varphi}\}$
\begin{equation}
\Psi_{2D}(1,2,3)
=\mu_0r_0^2\prod_{1(2)\leq j <k \leq 3}\sin(\varphi_{jk}/2)
\end{equation}
where $\mu_0$ is a constant.
The product lower bound is 2 if the positions of the particles 
are  as in Fig.~\ref{fig:2dho}a, while it is 1 if the positions  
are as in Fig.~\ref{fig:2dho}b.
Note that in the case of configuration in Fig.~\ref{fig:2dho}a 
rotation by $\pi$
flips the wave function sign so that the wave function vanishes
while for the configuration in  Fig.~\ref{fig:2dho}b it does not.
 For six particles in the positions outlined in  Fig.~\ref{fig:2dho}d the
wave function is clearly nonvanishing since  we can write 
\begin{equation}
\Psi_{2D}(1,...,6)
=\mu_0(r_2^2-r_1^2)r_2^6\prod_{j<k}^6\sin(\varphi_{jk}/2)
\end{equation}
where $\mu_0$ is some constant (in Fig.~\ref{fig:2dho}d we have
 $r_1=0$).
Alternatively,
the six particles can be positioned in pairs as sketched 
 in Fig.~\ref{fig:2dho}c where the particles are positioned 
on the HF node so that the noninteracting wave function vanishes.

For the sake of completeness we provide the expression for the wave function for
a general system
with size $M$  
\begin{equation}
\Psi_{2D}(1,...,N) =\nonumber
\end{equation}
\begin{equation}
=\mu(r_1, ..., r_{\tilde M})\prod_{k=1}^{\tilde M}\left[
\Psi_{1D}(I_k)\prod_{1\leq j<k}(r^2_k-r^2_j)^{n_j}\right]
\label{eq:2dho}
\end{equation}
where $\tilde M= {\rm int}[(M+1)/2]$ is the integer part of $(M+1)/2$. The particles with
indices in the subset $I_k$ lie on a circle with the
radius $r_k$ and $n_k=2(M+1-2k)+1$.  The prefactor is again a nonnegative function
of  
 $r_j$ which has 
no impact on the nodes
and is constant for rotations. From the last equation we see that in cases when $M$ is even, eg,
$M=6, N=21$, we end up with factorization of the type $11\times 7 \times 3$ particles 
while for $M$ odd, eg, $M=7,N=28$, we get $13\times 9\times 5 \times 1$. That means that the last
three particles ($M$ even) or the last six particles ($M$ odd) can be always arranged into possibilities 
as sketched in
Fig~\ref{fig:2dho}a-d. 
It is now clear that we can prove that
the spin-up and -down configuration space are interconnected
for the cases when $J=2,3$ as specified above. 
Using the alternative configurations for three and six particles we can arrange 
the systems into such 
positions that the rotation by $\pi$ exchanges odd number of pairs in both spin channels.
For example, if $N$ is
odd and $(N-1)/2$ is even
we position three particles on a triangle as given in Fig.~\ref{fig:2dho}b. 
That takes out one pair from
the $(N-1)/2$ exchanges and the proof then follows using the arguments for $3D$ harmonic fermions.
Similarly, if  $N$ is even and $N/2$ is also even,
we position one particle of each spin at the origin and 
five from each spin channel on a circle. This eliminates three pairs from $N/2$ and again we end 
up with remaining number of pairs being odd so that the rest of the proof follows similarly to the
$3D$ case.

\subsection{VI.c. Spin unpolarized interacting $2D$ and $3D$ homogeneous electron gas.}

\begin{figure}[ht]
\centering
\includegraphics[width=1.5in,clip]{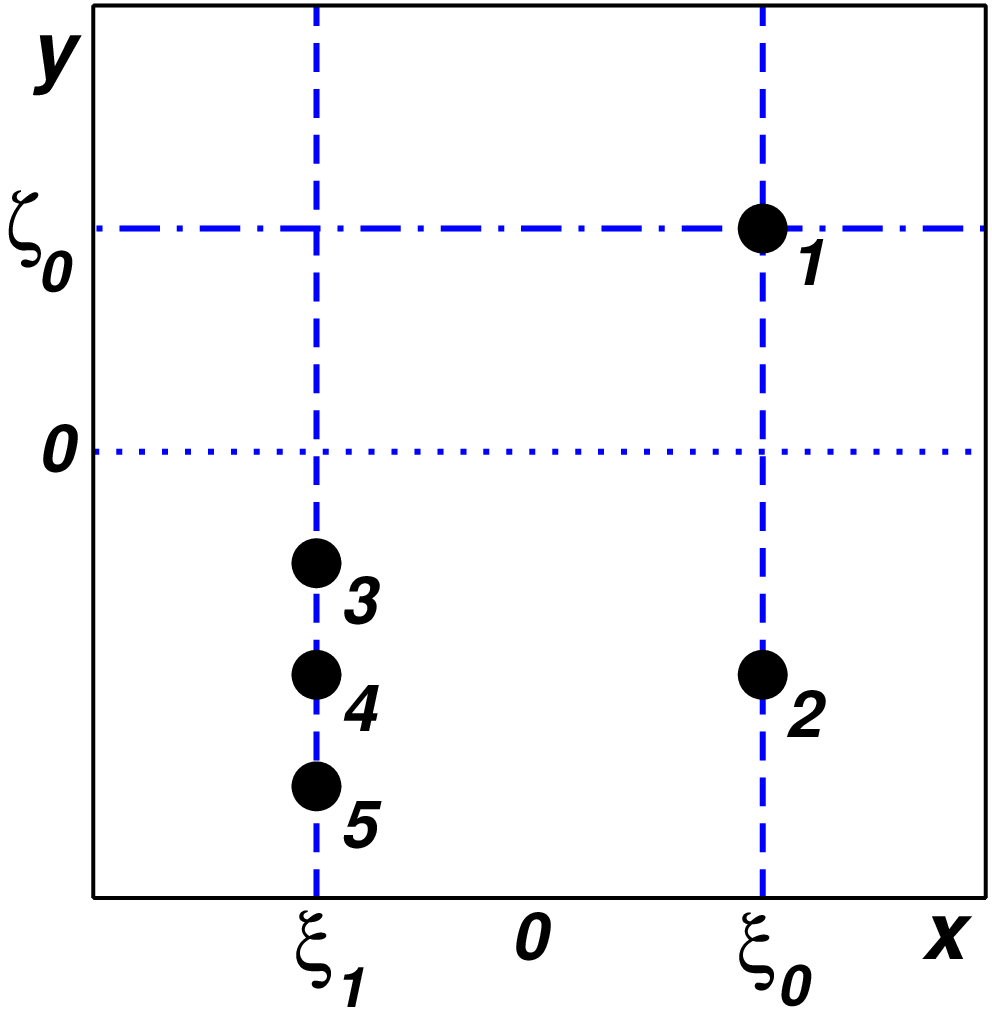}
\caption{Positions of five spin-up electrons from an interacting ten-fermion system
in the $2D$ periodic box $(-\pi,\pi)^2$.
For clarity, the five down-spin particles  
are not shown. The spin-down particles are positioned in the same pattern
on two lines $x=const$.}
\label{fig:2dhegunpol}
\end{figure}

Let us now prove the two nodal cells for the spin-unpolarized closed shells
for $d>1$ homogeneous  gas with interactions. We use the relevant definitions
introduced  previously for the spin-polarized periodic fermion gas (section III.a.).
We assume a $2D$ system of $2N$ particles in the spin singlet ground state
where the one-particle states occupy the Bloch states within the Fermi disk with the
radius 
 $k_F$. Let us first consider $2N=10$ particle 
case and we position the particles 
as in Fig.  \ref{fig:2dhegunpol}. 
The wave function is translationally invariant therefore the action of
$T^{y}_{\pi}$ leaves the wave function unchanged. 
 Assuming first that there is no interaction, we see that the
 translation  exchanges the particles $1,2$ but due to the invariance 
the wave function value does not change: that is possible only if the particles
are sitting on the node.
This also agrees with expressions for the wave functions derived before  
(see Eq. \ref{eq:five2dxi}.) Consider now that we switch-on interactions
and describe the particles by the BCS wave function as given above. 
The pairing orbital for the noninteracting gas is given by
\begin{equation}
\Phi_0(i,j)=\sum_{|{\bf k}|\leq k_F} \phi_{\bf k}(i)\phi_{\bf -k}(j)=1+2[\cos(x_{ij})+\cos(y_{ij})]
\end{equation}
where we used the occupied orbitals $\{1,e^{\pm ix},e^{\pm iy}\}$.
The correlated part is given by
\begin{equation}
\Phi_{corr}(i,j)=
\gamma\sum_{|{\bf k}|> k_F}\phi_{\bf k}(i)\phi_{\bf -k}(j)=
\nonumber
\end{equation}
\begin{equation}
\gamma[\cos(x_{ij})\cos(y_{ij})+\cos(2x_{ij})+\cos(2y_{ij})]
\label{eq:2dcorr}
\end{equation}
where $\Phi_{corr}(i,j)$ includes the next two unoccupied stars of states with $k=\sqrt{2},2$ 
and $\gamma$ is a variational parameter. 

Unlike for the noninteracting wave function (ie, $\gamma=0$) 
the BCS correlated wave function does not vanish and this is indeed the case due to the 
arguments presented above. The Slater 
matrix of uncorrelated wave function has linear dependency while for the correlated case
there always exists a configuration of lines with nonvanishing wave function. This is
due to the elimination of the linear dependency through addition of
virtual orbitals, as explained
for the $3D$ harmonic oscillator.
Note that as soon as the wave function
is translationally invariant the wave function does not vanish
for the whole translation path, implying that the spin-up and -down
nodal domains are interconnected and the wave function has only two nodal cells. 
It is straightforward to extend the proof
to arbitrary size closed-shell system. 
The configuration of particles can be given as follows: an odd number of pairs (eg, one) in each
spin channel
is positioned in a similar way as in
\ref{fig:2dhegunpol} so that the translation by $\pi$ exchanges the particles in the pair. For example,
for a spin-up pair 
$i,i+1$ we specify the coordinates as $y_i=-y_{i+1}=\pi/2$  and $x_i=x_{i+1}=\xi_0$ where $\xi_0$ is
distinct from $x$ coordinates of the rest of particles with the same spin.
The rest of the particles is positioned in such a way so that the $T_{\pi}^x$ brings them into
a symmetric position regarding the reflection around axis $y$. 
The wave function is translationally invariant and therefore uncorrelated wave function
vanishes while for BCS case it is, in general, nonzero. 
The two nodal cells property then 
follows using the same general arguments as in the previous cases.

\begin{figure}[ht]
\centering
\begin{tabular}{cc}
\includegraphics[width=2.6in,clip]{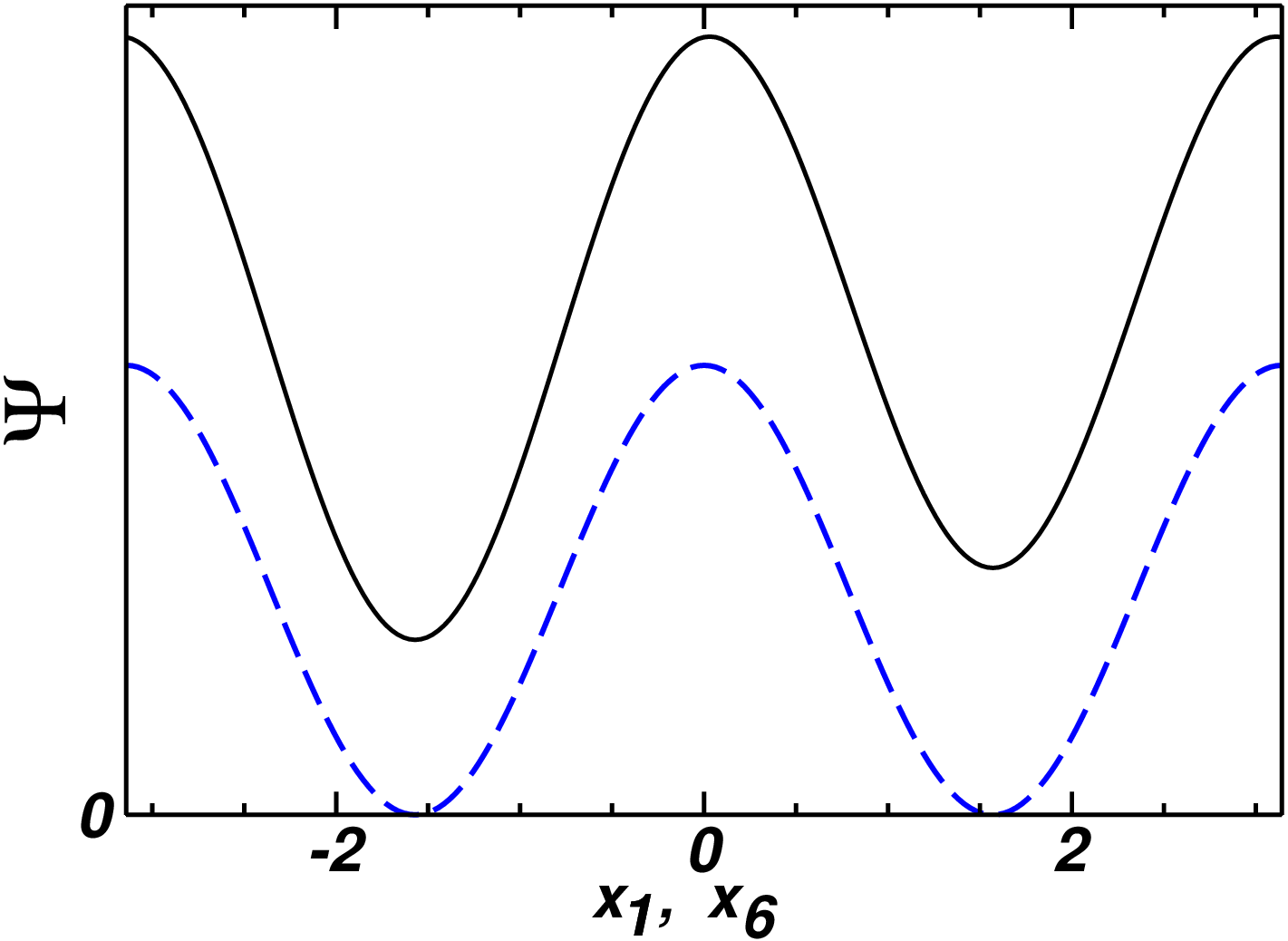} \\
(a)\\
 \quad \\
 \quad \\
 \quad \\

\includegraphics[width=2.6in,clip]{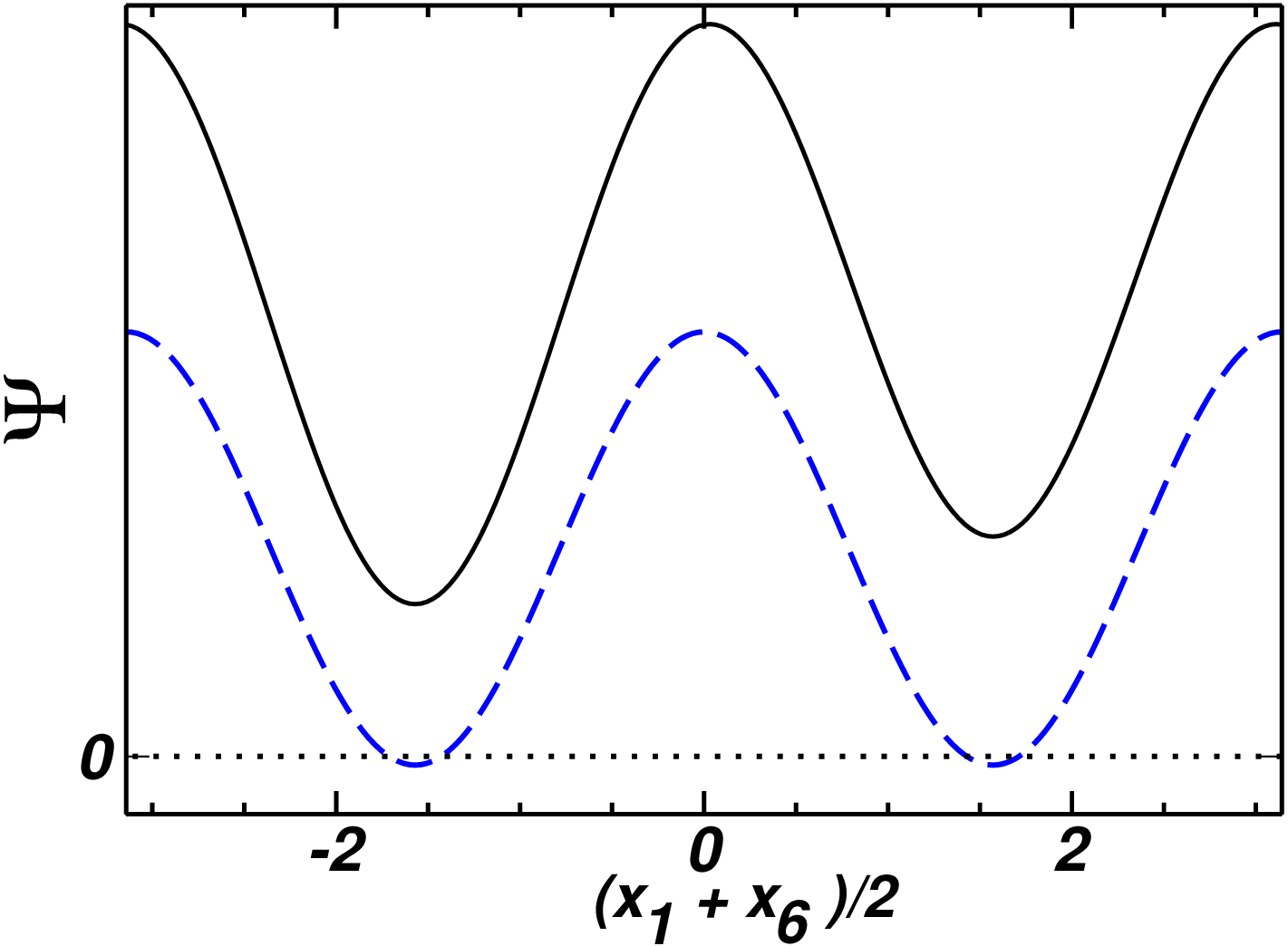} \\
(b) \\
 \quad \\
\end{tabular}
\caption{ Winding the particles 1 (spin-up) and 6 (spin-down) around the $2D$ box along 
the line $y=\zeta_0$. The dashed line is the Hartree-Fock wave function (arb.u.) while 
the full line is
the correlated BCS wave function.
a) Particles 1 and 6, 2 and 7, etc, share the same position. The uncorrelated wave function
touches the node quadratically since the spin-up and -down determinants are 
identical. b)  The positions of particles in spin-down channel are offset by 0.2
from the positions in the particle spin-up channel. 
The uncorrelated wave function
crosses the node multiple times since the spin-up and -down determinants are different.
}
\label{fig:wind}
\end{figure}

The correlated wave function has another important property  namely that one can wind around the
periodic box a singlet pair of particles (ie, spin-up and spin-down pair) without hitting
the node. Indeed, this is implied by the fact that there are only two nodal cells: 
once this is the case one can then always
find such a path that the pair of particles can wind around the box without
node-crossing. For example, for the ten-particle 
example above we can wind spin-up and -down particle pair around along $y=\zeta_0$ without hitting
the node. We plot the wave function for two types of particle positions: first, the
 spin-down particles are at identical 
positions as the spin-down ones, ie, ${\bf r}_6={\bf r}_1$, ${\bf r}_7={\bf r}_2$, etc 
(Fig.~\ref{fig:wind}a);
second, the particles  in spin-up and -down channel were offset by small displacement
${\bf r}_{i+5}={\bf r}_i+0.2,$ $i=1,...,5$, see 
 Fig.~\ref{fig:wind}b. 
Consider now the simultaneous translation $T_{2\pi}^y$ of the particles 1 and 6. This 
translation winds
the pair around the box and the wave function values for the paths are  
plotted in Fig.~\ref{fig:wind}a,b. 
The value of the displacement between the particle positions neither the repositioning
of the particles around the box is not crucial and there 
exists a significant size subspace of particle positions for which the winding of the pair
is possible.
We have chosen just two simple examples as qualitative illustrations.

Note that the wave function enables to wind also a single particle without the node crossing. 
However, single particle winding is possible both for correlated and uncorrelated wave functions
since, say, the spin-up HF part has two nodal domains and correlated wave function 
only smoothes out the HF nodes. Therefore this property is
not affected by the correlation in any significant manner.

The effect that singlet pair of particles can pass 
through nodal openings between the spin-up and -down subspaces
have been demonstrated for small number of particles before,
for example, in our paper on employing pfaffian pairing functions for electronic structure
problems \cite{bajdichpfaffians}. 
The case here illustrates that this property stems from interconnected
spin-up and -down subspaces and therefore applies to similar two nodal cell wave functions,
in general.

It is well known that a Fermi liquid, such as the homogeneous electron gas, becomes unstable to
a weak attractive interaction, develops Cooper pair instability and opens the superconductivity
gap \cite{supercond}. This effect is captured by the BCS wave function which we used in the proof above. 
Cooper pairs can therefore wind around the box without hitting the node although this effect
is not exclusive to Cooper pairs only. 
 In fact, the connectedness of  spin-up and -down
subspaces is a rather generic property in the sense 
that it appears also in systems which are not necessarily superconducting, ie, in Fermi liquids
with repulsive interactions. Since superconductivity is characterized by macroscopic phase
coherence and a number of other properties, which might or might not be related to
the nodal topologies, on the basis of the analysis of the BCS wave function above
one expects  that the  nodal opening 
{\it necessarily} appears in the 
superconducting state, 
however,
for the superconductivity to occur this condition is not sufficient.
In addition, here we assume only
the simplest $s-$wave pairing; for the $p-$wave or higher angular momentum pairing the situation
is less clear and needs to be further investigated. 

\section{\label{sec:dm} VII. Nodes of fermionic density matrices.}
 
In this part we generalize the ideas presented in preceding 
section to temperature dependent density matrices \cite{davidnode}. Consider
first a system of spin-polarized fermions. The temperature/imaginary time density matrix  
is given by
\begin{equation}
\varrho(R',R,\beta)= \sum_{n} e^{-\beta E_n} \Psi^{*}_n(R')\Psi_n(R)
\label{eq:dm1}
\end{equation}
where $\beta$ is the inverse temperature and  
the sum is over the complete system of eigenstates of a given Hamiltonian $H$. 
It is clear that the density matrix
is antisymmetric in particle exchanges in the same manner as  
the wave function $\Psi(R)$ or $\Psi(R')$. Therefore the notion of fermion
nodes can be generalized also to density matrix 
in $(2dN+1)$ dimensions since there is an explicit dependence on $\beta$ as well. As pointed
out elsewhere \cite{davidnode}, for fixed $R'$ and $\beta$ one can study the nodes and nodal cells 
in the $dN$ dimensional $R-$subspace. Similarly to the wave function, the node then becomes
$(dN-1)$-dimensional manifold with the generalization
that it is dependent on $R'$ and $\beta$. 
For the fixed  $R'$ and $\beta$
the tiling property holds in the same
manner as for the wave functions. The key additional feature of the density matrix is
that once there are only two nodal cells at some initial $\beta_0$ than this property holds for 
any $\beta >\beta_0$ \cite{davidnode}. This is not difficult to understand since the density matrix fulfills
the following linear equation
\begin{equation}
-{\partial \varrho(R,R',\beta)
\over \partial \beta}=H\varrho(R,R',\beta)
\label{eq:dm2}
\end{equation}
with an initial condition
\begin{equation}
 \varrho(R,R',0) = {\cal A} \delta(R-R') ={\det }[\delta({\bf r}_i-{\bf r'}_j)]
\label{eq:dm3}
\end{equation}
where ${\cal A}$ is the antisymmetrizing operator.

We now understand that for almost any  Hamiltonian with interactions 
the density matrix will have only
two nodal cells for sufficiently large $\beta$ (ie, at  low temperatures). This is due to the fact
that at sufficiently low temperature  the ground state becomes dominant
(Eq. \ref{eq:dm1}). The key point now is  to show that this
is the case also for {\it high} temperatures.  
The free particle density matrix 
is given by
\begin{equation}
 \varrho (R,R',\beta) = (2\pi\beta)^{-dN/2}{\rm det} \left[\exp(-|{\bf r}_i-{\bf r'}_j|^2/2\beta)\right]
\label{eq:dm4}
\end{equation}
  where we assume atomic units with $\hbar^2/m=1$.
 (For other than free
  boundary conditions, such as for the periodic ones,
 the expression has to be modified accordingly.)
Note that this
 density matrix is universal since   
at sufficiently high
temperatures the interactions become irrelevant. 

There are several ways how to prove that 
the high-temperature density matrix has only two nodal cells.
For very small $\beta$ one can use the induction as follows. 
Assume that $N$ particles are described by the density matrix given by
Eq.~\ref{eq:dm4} and the particles are connected by triple exchanges for a fixed $R',\beta$.
We add an additional particle 
 with the label $N+1$ to the system which 
occupies a certain region of the configuration (free) space.
The particle $N+1$ is positioned at the border of the occupied region
and let us assume that the particles with labels 
$N-1$ and $N$ are its closest neighbours. Let us move the three particles $N-1, N, N+1$
 away from the rest without crossing the node (what can be always done
by appropriate positioning). 
Since for small $\beta$ the overlaps of gaussians become small, one can factorize the determinant 
into a product of the three particles $N-1,N,N+1$ determinant and the determinant for the
 rest. 
For the three particles the density matrix has only two nodal cells as one can show easily
\cite{davidnode}
and therefore
the additional particle is connected. This applies to both $R$ and $R'$ subspaces since they must
have identical properties. (In what follows we will show that, in fact, at high temperatures 
the primed and unprimed spaces are connected as well.)  

There is also an alternative proof which is interesting also on its own since it provides
a different view on the density matrices. 
Note that the functional form of the free particle density matrix 
has a unique property. Comparing the Eq.~\ref{eq:dm4} with the  
 BCS wave function (Eq.~\ref{eq:bcs}) we see that the
high-temperature density matrix can be identified with a BCS wave function
if we properly define an underlying effective model. Instead of
our original  system of $N$ spin-polarized fermions, let as consider
a model system with $2N$ particles so that
 the configurations $R$ and $R'$ denote positions
of these {\it different } sets of particles, which we will call for simplicity unprimed and 
primed particles. Let us then define a new Hamiltonian ${\tilde H}(R,R')$ 
with
an effective quadratic interaction between the  unprimed and primed particles as
given by
\begin{equation}
{\tilde H}(R,R') =  T + T' + V_0\beta^{-1} \sum_{i,j} |{\bf r}_i - {\bf r'}_j|^2
\label{eq:dm5}
\end{equation} 
where $T$ and $T'$ denote kinetic energy operators for the corresponding sets of particles and 
$V_0$ is a constant with appropriate dimensions. 
Note that particles within the given set, say, unprimed, are antisymmetric but 
otherwise
 do not interact with each other. The interaction appears only between the primed 
and unprimed degrees of freedom as given by the Hamiltonian $\tilde H(R,R')$.
For $\beta \to 0$ the exact wave function for this system is a BCS wave function 
$\Psi(R,R')={\rm det}[\phi({\bf r}_i,{\bf r'}_j)]$ where the indices  
$i$ and $j$ label unprimed and primed particles, respectively. 
The pairing function $\phi({\bf r}_i,{\bf r'}_j)$ is obviously the gaussian given above. 
It is not too difficult to demonstrate that this wave function
(and the density matrix) has only two nodal cells. 
For example, we can expand the gaussian into plane waves
\begin{equation}
\exp(-|{\bf r}_i-{\bf r'}_j|^2/2\beta)=\sum_{\bf k}c_{\bf k}e^{i{\bf k}\cdot({\bf r}_i-{\bf r'}_j)}
\label{eq:dm6}
\end{equation}
where $\{c_{\bf k}\}$ are expansion coefficients.
 The sum is over states within the Fermi sphere of
a periodic box which accommodates $2N$ particles for a given density. Let us specify that
$N$ is large and $\beta$ is such that 
the system can be considered classical so that the actual interaction in the 
original Hamiltonian $H$ is irrelevant.
Then the density matrix corresponds to the Hartree-Fock product
\begin{equation}
{\rm det}[e^{i{\bf k}_i\cdot {\bf r}_j}]{\rm det}[e^{-i{\bf k}_i\cdot {\bf r'}_j}]
\label{eq:dm7}
\end{equation}
However, we have already proved that such system has only two nodal cells. In addition, 
if the sum includes also "excited states"
(ie, beyond the Fermi sphere) due to the primed-unprimed interactions, 
 we find that  the unprimed and primed 
nodal cells are interconnected.  Therefore at classical temperature  $\beta_0$
the density matrix has only two nodal cells and then the same is true for arbitrary
$\beta>\beta_0$. This primed-unprimed
interconnection becomes less pronounced and ceases to exist at $\beta \to \infty$ since
then the density matrix is proportional to the "noninteracting" product
$\Psi_0(R)\Psi_0(R')$ where $\Psi_0$ is the ground state for the original physical system 
of $N$ fermions described by $H$. This proof is therefore based on an interesting duality between the 
$N$ spin-polarized fermions at classical temperatures and the model
system with $2N$ particles with a temperature-dependent,
 harmonic interaction between the unprimed and primed
subspace particles. 

Possibly, there might be also a third way how to prove the minimal number of nodal cells for
the high-temperature density matrix  
through diagonalization of the quadratic Hamiltonian $\tilde H$ (we have not 
investigated this  possibility). 
The fact that also the density matrices have two nodal cells is important for the path 
integral Monte Carlo methods which for fermions often employ the fixed-node 
approximation adapted for the 
path integrals \cite{davidhe}. 

\section{VIII. Discussion and conclusions.}
 
Inspired by previous
conjectures and numerical studies
\cite{jbanderson,davidnode,lesternode,sorella,foulkes,dariobe,bajdichnode}
the presented analysis and proofs generalize and clarify the properties
of ground states fermion nodes. 
We have employed symmetries, wave function factorizations
and triple exchanges 
to prove that for $d>1$ the closed-shell ground states 
have two nodal cells for arbitrary number of particles
in several paradigmatic models.
In this paper the proofs were carried for closed shells, mainly to avoid 
additional complications from degeneracies and some
of these aspects will be addressed in subsequent papers.
   
It is perhaps more interesting
to discuss viable possibilities when the ground state wave functions might have 
{\it more 
than two} nodal cells. Clearly, it is easy to generate more nodal cells by imposing 
additional symmetry
or boundary conditions. Another possibility comes from very strong interactions,
for example, whenever the effect of interaction
would
 become competitive with the kinetic energy increase from forming additional 
nodes it is possible that more than the minimal two nodal cells can form. 
One can also construct nonlocal interactions
which would violate some of the properties mentioned here (eg, the tiling property),
or reorder the states so that, for example, excited states could 
lie below the ground state, etc.
Another candidate for unusual effects in the nodal structure are open-shell systems with
large degeneracies and densities of states at the Fermi level.  Some of these 
interesting issues will be subject of future studies.

In conclusion,  we have presented a number of new results which reveal the structure
of nodes and nodal cells
 of fermionic wave functions. 
Building upon ideas introduced in previous paper
we were able to demonstrate the minimal number of two nodal cells in several 
spin-polarized models such
as noninteracting fermions in a periodic box, in a box with zero boundary conditions,  fermions
on a spherical surface, etc. This enabled us to formulate a theorem
which states that in $d>1$ the minimal two nodal cells are present for any
Slater determinant with monomial matrix elements for any size which
allows for a closed-shell nondegenerate ground state. We have shown that this 
property extends also to cases which cannot be described by the Slater matrix of monomials
such as the noninteracting and HF atomic states up to the $3d$ shell.
We have studied the effect of interactions on the noninteracting nodal cells
of spin-unpolarized, closed-shell singlets.  Our results show
that, in general, the interactions smooth out
the noninteracting multiple nodal cells into the minimal number of two. 
For interacting homogeneous
electron gas we have demonstrated that the two nodal cells allow 
singlet pairs of particles to wind around 
the periodic box without crossing the node.
Finally, we have 
shown that the temperature/imaginary time
 density matrix has very similar nodal structure and therefore
the minimal two nodal cells property applies also to density matrices with
important implications for path integral Monte Carlo 
simulations. We have demonstrated this by using an appropriate mapping of the density matrix
onto the ground state of a model systems with twice as many particles interacting 
with temeperature-dependent harmonic potentials.


I am grateful to J. Koloren\v c, M. Bajdich and L.K. Wagner for reading the manuscript, 
comments and discussions.  
I would like to acknowledge the support by
NSF DMR-0102668, DMR-0121361 and  EAR-0530110 grants.

\end{document}